# Structural instability of the rutile compounds and its relevance to the metal–insulator transition of VO$_2$

Zenji Hiroi

*Institute for Solid State Physics, University of Tokyo, Kashiwa, Chiba 277-8581, Japan*

ABSTRACT

The metal–insulator transition (MIT) of VO$_2$ is discussed with particular emphasis on the structural instability of the rutile compounds toward dimerization. Ti substitution experiments reveal that the MIT is robust up to 20% Ti substitutions and occurs even in extremely thin V-rich lamellas in spinodally decomposed TiO$_2$–VO$_2$ composites, indicating that the MIT is insensitive to hole doping and essentially takes on a local character. These observations suggest that either electron correlation in the Mott–Hubbard sense or Peierls (Fermi-surface) instability plays a minor role on the MIT. Through a broad perspective of crystal chemistry on the rutile-related compounds, it is noted that VO$_2$ and another MIT compound NbO$_2$ in the family eventually lie just near the borderline between the two structural groups with the regular rutile structure and the distorted structures characterized by the formation of dimers with direct metal–metal bonding. It is also shown that the two compounds of the rutile form do not follow the general trends in structure observed for the other rutile compounds, giving clear evidence of an inherent structural instability present in the two compounds. The MITs of VO$_2$ and NbO$_2$ are natural consequences of structural transitions between the two groups, as all the $d$ electrons are trapped in the bonding molecular orbitals of dimers at low temperatures. Such dimer crystals are ubiquitously found in early transition metal compounds having chain-like structures, such as MoBr$_3$, NbCl$_4$, Ti$_4$O$_7$, and V$_4$O$_7$, the latter two of which also exhibit MITs probably of the same origin. In a broader sense, the dimer crystal is a kind of "molecular orbital crystals" in which virtual molecules made of transition metal atoms with partially-filled $t_{2g}$ shells, such as dimers, trimers or larger ones, are generated by metal–metal bonding and are embedded into edge- or face-sharing octahedron networks of various kinds. The molecular orbital crystallization opens a natural route to stabilization of unpaired $t_{2g}$ electrons in crystals.

# 1. Introduction

## 1.1. Motivation and strategy

The metal–insulator transition (MIT) is one of the most dramatic phenomena observed in condensed matter science, in which a high-temperature metallic state is transformed into a low-temperature insulating state at a critical temperature ($T_{MI}$) [1-5]. It can occur purely electronic in origin, such as a Mott transition from a correlated metal to a Mott–Hubbard insulator driven by Coulomb repulsions and a charge-density wave transition or a Peierls transition associated with certain Fermi-surface (FS) instability. The former does not require crystal symmetry breaking, while, in the latter, a structural transition that removes the FS instability is always accompanied. MITs observed in various transition metal compounds have been focused as candidates for the Mott transition [3]. However, most of them, possibly except for the high-temperature transition of $V_2O_3$, are accompanied by structural changes, which makes it difficult to identify the Mott mechanism as the major cause and to extract the essential feature of the Mott transition experimentally. An electronic MIT and a structural transition often take place simultaneously, because electrons that are going to localize owing to Coulomb repulsions may feel more electron–phonon interactions or those to be captured by a lattice may feel more Coulomb repulsions [4]. As a result, actual phenomena at MITs are always complex and difficult to interpret within a simple scenario; to specify the mechanism of an MIT has been proved troublesome and it often remains a chicken-and-egg riddle between two or more factors.

The MIT of vanadium dioxide $VO_2$ in the rutile (R) family of compounds is one of the long-standing issues left in the research of transition metal oxides. It has been extensively studied by many researchers in the field for more than half century since the first report by Morin in 1959 [1] and revisited from time to time. However, as is usually the case, the mechanism still remains controversial, penduluming among the Mott–Hubbard type and the Peierls (band) type; there has been a "tug of war" between electron correlation and band sects [6,7].

In addition to the academic interest, the MIT of $VO_2$ is also important for technological applications such as sensors and switching devices, because abrupt changes in resistivity and optical absorption at the transition at $T_{MI}$ = 340 K are useful for them [8,9]. For example, an infrared imaging device operating at room temperature has been produced, in which absorbed infrared radiation changes the temperature of a microbolometer made of $VO_2$ with high sensitivity due to a large temperature coefficient of resistance at the transition [10]. Recently, an active control of the $T_{MI}$ has been carried out in thin films or beams by means of electric field or strain [11,12]. In order to improve the performance of any device, a basic understanding of the microscopic mechanism of the MIT is necessary.

Recently, we have studied the $TiO_2$–$VO_2$ system and obtained remarkable results [13], as will be mentioned later in detail. Although complete solid solutions (SSs) $V_{1-x}Ti_xO_2$ could be prepared by quenching from high temperature, the SSs with intermediate compositions are thermodynamically unstable below 830 K and decompose into V-rich and Ti-rich phases by the spinodal decomposition mechanism: there is a miscibility gap in the phase diagram of the $TiO_2$–$VO_2$ system. Surprisingly, a clear and sharp MIT was observed in very thin lamellas of the V-rich phase containing 10–20% Ti at slightly reduced transition temperatures. Note that the $Ti^{4+}$ substitution simply brings 'holes' into the $d$ bands. This observation has led me to consider that a structural component is primary important rather than electron correlations or band instabilities. In this paper, I will review the crystal chemistry of all rutile-related compounds in order to get insight on the true mechanism of the MIT of $VO_2$.

I discuss here the origin of the MIT of $VO_2$ in terms of a structural instability that is underlying in the rutile structure. What I try to show is that $VO_2$, as well as another MIT compound $NbO_2$ in the rutile family, lie very close to the boundary between the rutile and the distorted structures comprising dimers of transition metal atoms; the tetragonal axial ratio $c/a$ of the rutile structure can be a crucial parameter and the boundary lies at $c/a$ = 0.625. I consider that the MIT is nothing to do with electronic instabilities caused by electron correlations or the Peierls (FS) instability but must be driven by a structural instability between the two structural polymorphs, which is characterized by the formation of a local metal–metal bonding, emphasized long time ago by pioneering solid state chemists like Magnéli and Goodenough [14,15] but has been unregarded by most physicists.

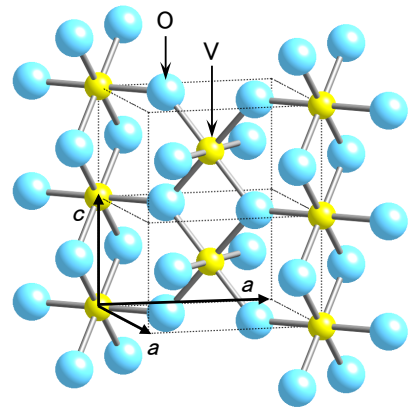

Fig. 1. Crystal structure of the rutile type. It contains strands made of edge-sharing $VO_6$ octahedra running along the $c$ axis. Two strands at the corner ($z$ = 0) and center ($z$ = 1/2) of the basal square are rotated by 90° and connected to each other via their common vertices.

## 1.2. MITs in $VO_2$ and related compounds

The MIT of $VO_2$ is dramatic: a large discontinuity of several orders of magnitude in resistivity and a sudden jump in magnetic susceptibility are observed across the transition at $T_{MI}$ = 340 K upon heating and at a slightly lower temperature by several kelvin upon cooling [1,15,16]. It is a first-order transition from a metallic paramagnet to an insulating diamagnet upon cooling. A structural transition takes place simultaneously from the high-temperature rutile structure (Fig. 1) to the low-temperature monoclinic structure called M1, which is characterized by the formation of V–V pairs slightly tilted from the $c$ axis [17-19].

Goodenough proposed a phenomenological model that appeared to rationalize many of the properties [18,20,21]. According to his model and the following study by others, the fundamental electronic structure of the rutile-type transition metal oxides is composed of two states originating from the $t_{2g}$ orbitals in the octahedral crystal field: the $d_{//}$ orbitals spreading along the $c$ axis are nearly nonbonding and the $\pi^*$ orbitals strongly hybridizing with the O $2p$ $\pi$ state are isotropic. Both states are partially occupied in the metallic state of $VO_2$. In the M1 phase of $VO_2$, the pairing of V atoms causes a bonding–antibonding splitting of the $d_{//}$ band, with the lower $d_{//}$ band fully occupied and the upper $d_{//}$ and $\pi^*$ bands completely vacant, opening an energy gap of ~0.7 eV.

On the basis of a large accumulation of experimental and theoretical studies, the mechanism of the MIT has been greatly debated for decades from various perspectives. In the early days, Marinder [14] and Goodenough [18,20,21] pointed out that the formation of homopolar V–V bonding could account for the MIT. Later on, cooperative behaviors of $3d$ electrons have been emphasized: two major scenarios assume the Peierls and Mott–Hubbard mechanisms. Since such a dimerization as in the M1 structure is characteristic of the Peierls transition for a one-dimensional electronic system, there is a reason to believe the former scenario [22,23]. On the other hand, the latter mechanism is suggested mainly because the band structure of the insulating phase can not be quantitatively explained only by the structural deformation but by incorporating electron correlations for V $3d$ electrons [24,25]. More recently, a sort of compromise mechanism called the Mott–Peierls scenario has been discussed [26,27]. For example, a correlation-assisted Peierls transition was proposed, in which a Peierls gap opens by the formation of dynamical V–V singlet pairs due to a strong Coulomb correlation [26]. On one hand, an orbital-assisted Peierls transition is expected to occur by massive orbital switching when the system lies close to a Mott insulating regime [27]. Although a general consensus may have not yet been reached, one can definitely say that the structural deformation plays an important role in the mechanism of the MIT of $VO_2$.

There is one more MIT compound $NbO_2$ in the rutile family [28-31]. It has been less studied compared with $VO_2$, mostly because its transition temperature is too high, $T_{MI}$ =1070 K, to carry out various experiments. The basic mechanism of the MIT of $NbO_2$ must be identical to that of $VO_2$, because both of them have the $d^1$ configuration and also because similar structural transitions that are characterized by the formation of metal–metal pairs occur in them. The true mechanism should give a persuasive explanation for the two MITs.

In addition, there is another related family of compounds that exhibit MITs; the Magnéli phases, $Ti_nO_{2n-1}$ and $V_nO_{2n-1}$ [32-35], which exist between $TiO_2$ ($VO_2$) and $Ti_2O_3$ ($V_2O_3$) with mixed-valence metal ions. The structures of the Magnéli phases are obtained by periodically introducing a stacking fault with an extra plane of metal atoms (called a shear plane) into the mother rutile structure; the composition is decided by the periodicity. MITs are observed in most of them: $Ti_nO_{2n-1}$ ($n$ = 4, 5, 6) [34,36-38] and $V_nO_{2n-1}$ ($n$ = 4, 5, 6, 8, 9) [35,36,39-41]. Markedly, pairings of metal atoms commonly occur in their insulating states [37,40,41]. In spite of the obvious similarity in the crystal structures and phenomena at the MITs, the MITs of the Magnéli phases have been discussed separately from that of $VO_2$ by researchers except for Chakraverty who classified the MITs of $VO_2$, $NbO_2$ and $Ti_4O_7$ as pairing interaction transitions [42]. I believe that there must be a common origin behind all these MITs in the rutile-based compounds.

*1.3. Substitution effects in $VO_2$*

Among various experimental results on $VO_2$, cation substitution effects on the $T_{MI}$ and the crystal structures are quite important [18,43]. There are two classes of substitution systems $V_{1-x}M_xO_2$: $T_{MI}$ increases slightly for M = $Cr^{3+}$ [19,44-46], $Al^{3+}$ [47], $Fe^{3+}$ [48], and $Ti^{4+}$ [13,49-52], while it decreases rapidly for M = $Nb^{5+}$ [49,53], $Mo^{6+}$ [54], and $W^{6+}$ [55,56]. In terms of a simple electron counting with $n$ being the number of excess electrons per formula unit, a $Cr^{3+}$ ion gives one hole and two electrons so that $n = x$ in total, while a $Ti^{4+}$ ion introduces one hole; $n = -x$. A $Nb^{5+}$ ion gives one electron and a $Mo^{6+}$ or a $W^{6+}$ ion adds two electrons. Figure 2 compares the $n$ dependences of $T_{MI}$ for various substitutions. Apparently, a simple expectation for electron or hole doping into a Mott–Hubbard insulator is confounded. Especially, the fact that $T_{MI}$ increases with electron doping by the $Cr^{3+}$ substitution and hole doping by the $Ti^{4+}$ substitution seems difficult to interpret within the Mott–Hubbard scenario: most Mott–Hubbard insulators are rendered metallic by a little doping into a half-filled band as observed in $La_{1-x}Sr_xTiO_3$ or $La_{2-x}Sr_xCuO_4$ [5]. It also throws doubt on the Peierls mechanism which generally requires a half-filled band of the one-dimensional character.

On the other hand, the large reductions of $T_{MI}$ for heavy metals like Nb, Mo and W are along with either scenario, but have not been discussed quantitatively. Recently, Holman and coworkers carried out a complete study on the $V_{1-x}Mo_xO_2$ system over a wide range of $0 < x < 0.50$ [54]. They found that the MITs remain sharp for $x < 0.20$ with reduced $T_{MI}$s, e.g. 160 K at $x = 0.16$, and becomes obscured for higher doping. More recently, Shibuya and coworkers studied the W substitution over a wide range of $0 < x < 0.33$ in the form of thin film and found that the MIT is completely suppressed at $x = 0.07$, but, surprisingly, it reappears above $x = 0.095$ with the $T_{MI}$ increasing with substitution to above room temperature at $x = 0.30$ [56]. The obtained phase diagram has been discussed in terms of competition between the Peierls and Mott instabilities [57]. I will come back to this interesting observation at the end of this paper.

It is known that the substitutions of trivalent $3d$ elements introduce other insulating phases between the rutile (R) and M1 phases. For example, the Cr substitution introduces the monoclinic M2 phase and the triclinic T phase: the MIT occurs between the R and M2 phases [19,44-46]. The reason why the $T_{MI}$ increases slightly with the Cr substitution is that the R–M2 transition temperature increases accordingly. This is also the case for the Ti substitution [13,50,51]. On the other hand, the transition temperature to the M1 phase decreases with the Cr or Ti substitution, as shown in Fig. 2 for the Ti substitution. In the M2 phase, V atoms are split into two sets of chains parallel to the rutile $c$ axis. On one set of chains V atoms occur in pairs with alternating long and short separations, as in those of the M1 phase, while, on the other set of chains, the V–V atoms form a zigzag pattern but remain equally spaced [46]. Thus, the M2 structure is considered as an intermediate structure between the R and M1 structures. A true mechanism of the MIT in $VO_2$ should give reasonable explanations to these complex substitution effects.

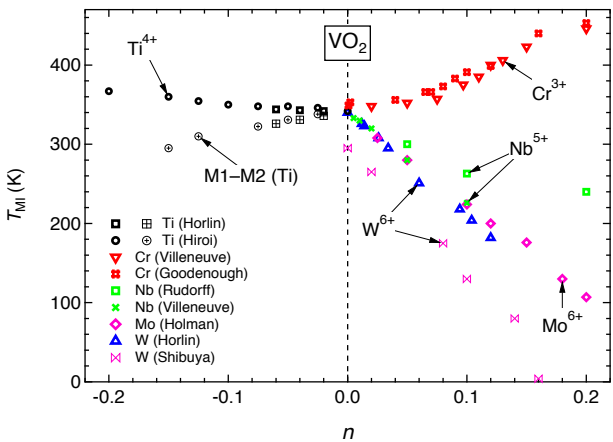

Fig. 2. Composition dependences of the MIT temperature $T_{MI}$ at various substitutions for V; $Ti^{4+}$ [13,50], $Cr^{3+}$ [44,45], $Nb^{5+}$ [49,53], $Mo^{6+}$ [54], and $W^{6+}$ [55,56]. The horizontal axis refers to the number of excess electrons per formula unit in the simple electron counting scheme: $n = -x$ (M = $Ti^{4+}$), $x$ ($Cr^{3+}$, $Nb^{5+}$), and $2x$ ($Mo^{6+}$, $W^{6+}$) in $V_{1-x}M_xO_2$. The M1–M2 transition temperature is also plotted for the Ti substitution.

## 2. Results and discussion

### 2.1. $TiO_2$–$VO_2$ system

#### 2.1.1. Spinodal decomposition

The Ti substitution for V is considered as the simplest case, because it does not bring extra $d$ electrons into the system as the others do but only "holes". The substitutions of most elements for V are limited below 20 mol%, above which either other phases are stabilized or conventional two-phase mixtures are obtained. In the Ti substitution system $V_{1-x}Ti_xO_2$, in contrast, it was reported that complete SSs could be prepared in polycrystalline samples or thin films [14,49,51,58]. In a later study on the single crystal growth, however, the presence of complete SSs was questioned because single crystals were obtained only for $x > 0.8$ even from nominal compositions of $x > 0.5$ [52]. On the other hand, in the V-rich side, an $x$–$T$ phase diagram including the R, M1, M2 and other phases is obtained [50,51], similar as in the case of $V_{1-x}Cr_xO_2$ [19,44–46]. Therefore, both the Ti- and V-rich sides of the phase diagram are well understood, but the middle region had remained less explored.

Very recently, we have shown that there is a miscibility gap in the $TiO_2$–$VO_2$ phase diagram [13]: an spinodal decomposition occurs in a similar manner as in the $TiO_2$–$SnO_2$ system [59] but below a much lower temperature of 830 K. The subsolidus miscibility gap of the $TiO_2$–$VO_2$ system is shown in Fig. 3. It would be helpful to describe how the microstructure develops with time duration after quenching from high temperature at a typical average composition of $x = 0.40$. A uniform SS with the rutile structure was obtained by rapid quenching from 1173 K into ice water. After being annealed at 673 K for 1 h, a compositionally modulated structure with a periodicity of $\lambda \sim 6$ nm was formed as a result of spinodal decomposition (SD), as schematically drawn in Figs. 3 and 4(b). Further annealing for 12 h at the same temperature gave a completely phase-separated (PS) microstructure that is composed of V-rich and Ti-rich lamellas alternately stacked along the $c$ axis; they have approximate compositions of $V_{0.86}Ti_{0.14}O_2$ and $V_{0.44}Ti_{0.56}O_2$ and thicknesses of 14 and 26 nm, respectively [Fig. 4(b)] [13].

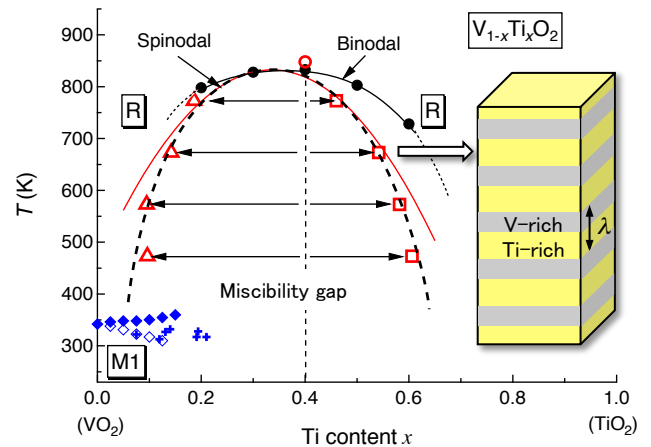

Fig. 3. Phase diagram of the $TiO_2$–$VO_2$ system [13]. A binodal line and a coherent spinodal line, which were determined by annealing experiments on the SS with $x = 0.4$ at four temperatures, are shown: triangles and squares represent the compositions of the V-rich and Ti-rich phases in the decomposed samples, respectively. The broken line represents a fit to the regular solution model: $T_s = 3330[1 - (x + 0.16)](x + 0.16)$. The critical temperatures of the R–M2 (solid diamond) and M2–M1 (open diamond) transitions observed in SSs are also shown in the left-bottom, together with those of the MITs (cross) observed in the V-rich phase in the PS samples.

#### 2.1.2. MIT

The MITs of the Ti-substituted samples have been probed by magnetic susceptibility $\chi$. A pure polycrystalline sample of $VO_2$ shows a sudden and large jump in $\chi$ at 342 K upon heating, as shown in Fig. 4(a), where a transition from a non-magnetic insulator to a paramagnetic metal and a structural transition from M1 to R take place simultaneously. Here the MIT temperature $T_{MI}$ is defined at the midpoint of the jump in $\chi$ upon heating; there is a thermal hysteresis, reflecting the first-order character, and a drop in χ occurs at a slightly lower temperature of 338 K upon cooling.

SSs show two-step transitions, as shown in Fig. 4(a) for $x$ = 0.1 and 0.2: the low- and high-temperature jumps correspond to the M1–M2 and M2–R transitions, respectively. The $x$ dependences of the transition temperatures are plotted in Figs. 2 and 3. The MIT occurs at the latter, as confirmed by resistivity measurements. Note that the MIT remains as sharp as in pure $VO_2$ even at $x$ = 0.2 (one-fifth of $d$ electrons are missing!) with keeping the first-order character, which indicates that the transition is robust against the Ti substitution, similar as in the Mo substitution [54].

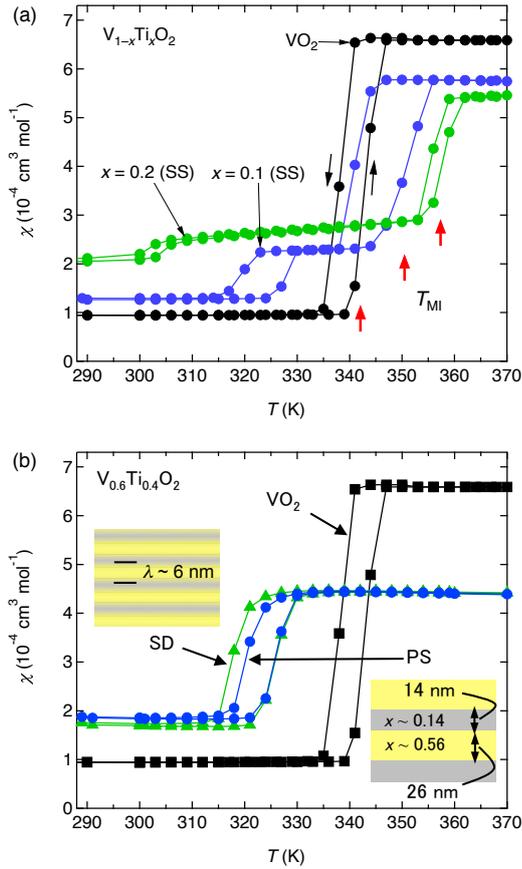

Fig. 4. (a) Magnetic susceptibility χ of polycrystalline $VO_2$ and solid solutions (SSs) $V_{1-x}Ti_xO_2$ with $x$ = 0.1 and 0.2 obtained by rapid quenching from 1173 K. A pair of curves represents two successive measurements upon heating and then cooling. Each SS shows two transitions with an MIT at a higher-temperature jump marked by a red arrow. (b) Evolution of χ for $V_{0.6}Ti_{0.4}O_2$ with the duration of annealing at 673 K. Sudden jumps are observed for the spinodally-decomposed (SD) and phase-separated (PS) samples at nearly the same temperatures of 325 K on heating. Their microstructures with composition modulations along the $c$ axis are schematically depicted in the inset [13].

MITs were not observed in SSs with $0.3 < x < 0.6$ but appeared after annealing at low temperatures, where spinodally modulated structures or phase-separated lamellar structures have been generated. An evolution of magnetic susceptibility with time duration for a series of polycrystalline samples with $x$ = 0.4 is shown in Fig. 4(b). A jump in χ emerges at 325 K for an SD sample with λ = 6 nm after annealing at 673 K for 1 h. A nearly same jump is observed at a nearly equal temperature in a PS sample further annealed for 12 h, which consists of 14-nm thick $V_{0.86}Ti_{0.14}O_2$ and 26-nm thick $V_{0.44}Ti_{0.56}O_2$ lamellas [13].

For reliable resistivity measurements, a single crystal with $x$ = 0.4 was grown by the floating-zone method and annealed in the same ways as done for polycrystalline samples. The grown crystal was easily cleaved along the (110) plane; a typical crystal is shown in the inset of Fig. 5(a). The results of resistivity measurements in Fig. 5(a) are consistent with those of magnetic susceptibility shown in Fig. 5(b), where no transition in the SS but in the PS crystal: semiconductor-like temperature dependences are observed in the SS crystal, while a clear jump at 322 K separates high-temperature metallic and low-temperature semiconducting behaviors in the PS crystal. Note that, in the SS crystal, there is little difference between two measurements with electrical currents running parallel and perpendicular to the $c$ axis, while a large anisotropy is observed in the PS crystal: the resistivity is about 1000 times larger at above $T_{MI}$ for current running along the $c$ axis, as semiconducting (more insulating) Ti-rich layers tend to prevent the current flow among the metallic (less insulating) V-rich layers above (below) $T_{MI}$. This dramatic change between the SS and PS crystals was perfectly reversible: annealing the PS crystal at 1173 K followed by rapid quenching to ice water removed the MIT completely.

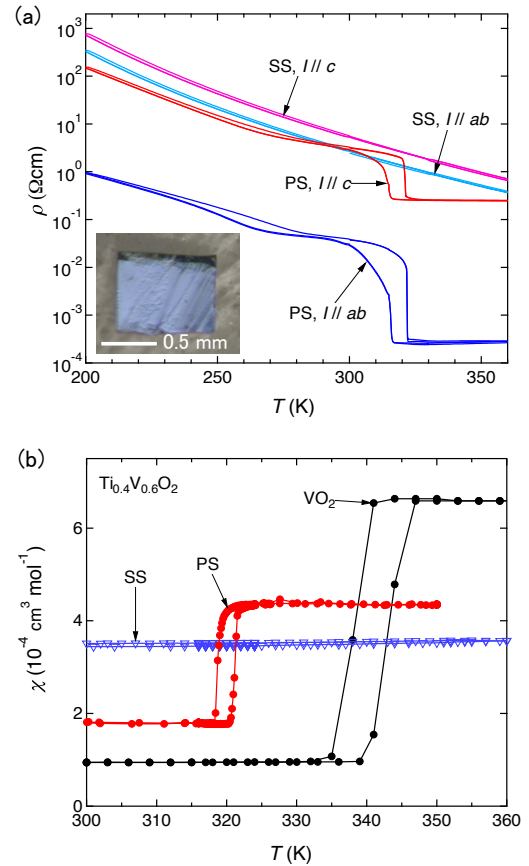

Fig. 5. Resistivity (a) and magnetic susceptibility (b) of single crystals of $Ti_{0.4}V_{0.6}O_2$. A typical crystal of 1 mm size is shown in the inset of (a), where the top surface corresponds to the (110)

plane. A solid solution (SS) crystal was obtained by annealing a pristine crystal at 1173 K and then quenching into ice water, and a phase-separated (PS) crystal was obtained by annealing the SS crystal at 673 K for 12 h. Resistivity measurements were carried out either with current parallel to the *c* axis or the *ab* plane for each crystal.

The presence of accompanying structural transitions has been confirmed by powder and single-crystal X-ray diffraction study, which will be reported elsewhere [60]. An M1–R structural transition takes place in the V-rich lamellas of the PS samples at the MIT, though the M2–R transition is expected in an SS with the corresponding composition of $x = 0.14$ (Fig. 2). This is consistent with the single transition observed in $\chi$ for the PS samples, compared with the double transitions for the SSs of $x = 0.10$ and $0.20$. The M2 phase may be unstable in the PS samples, probably because of strain due to lattice matching to the adjacent Ti-rich lamellas of the rutile structure. The reason for the reduction of $T_{MI}$ compared with pure $VO_2$ or the corresponding SS is also ascribed to an anisotropic strain from the adjacent Ti-rich layers [13,61-63].

The above results clearly indicate that the MIT is induced as a result of phase separation through which many of V atoms have condensed into the V-rich layers. Very important is that the MIT can occur as sharp as in pure $VO_2$ in the thin V-rich layers of 14 nm thickness in the PS sample or even thinner layers in the SD sample. Figure 6 schematically depicts a local structure expected for a V-rich layer in the SD sample. The thickness may be about 2 nm, including eight V atoms along the *c* axis with a few of them replaced by Ti atoms. As a result, the V chains are cut into short pieces. The fact that the MIT can occur in such an ultimate situation indicates that the MIT essentially takes on a local character, which may not be compatible with what one expects from the band picture or many-body effects arising from electron correlations.

We have thus obtained two important results in the $TiO_2$–$VO_2$ system. One is the robustness of the MIT against apparent hole doping in SSs: up to 20% Ti substitutions for V the $T_{MI}$ increases slightly and the transition remains as sharp as in pure $VO_2$. The other is the local character of the MIT which occurs even in extremely thin V-rich layers sandwiched by Ti-rich layers in an SD sample. I think that these two facts are difficult to understand within the previous scenarios.

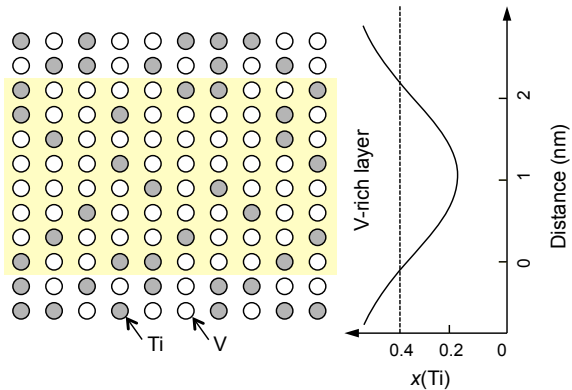

Fig. 6. Schematic drawing of a local structure expected for a V-rich layer in an SD sample with $\lambda = 6$ nm. In the 2-nm thick layer having such a composition modulation as depicted on the right, a few of eight V atoms are replaced by Ti atoms, resulting in segmented V chains, where a sharp MIT occurs.

*2.1.3. Orphan spins in the insulating phase*

The effects of substitutions on the magnetic properties of $VO_2$ have been studied to examine the nature of the insulating phase. Villeneuve and coworkers carried out complete study on the Nb substitution and observed a large Curie contribution in magnetic susceptibility at low temperatures, which corresponds to spin 1 per impurity with the Landé factor $g = 2$ in the dilute limit [53,64,65]. The impurity effects are readily understood in terms of the homopolar bonding picture of Goodenough assuming $V^{3+}$–$Nb^{5+}$ 'molecules' instead of $V^{4+}$–$Nb^{4+}$ ones. Compared with the Nb substitution, the Ti substitution is more straightforward as Ti exists definitely as $Ti^{4+}$ and brings one hole into the system. In fact, Hörlin and coworkers observed a similar Curie contribution corresponding to spin 1/2 per impurity for small Ti contents, as expected for the formation of $Ti^{4+}$–$V^{4+}$ 'molecules' [51].

We have obtained essentially the same results in our Ti substitution experiments as Hörlin and coworkers did [66]. Figure 7(a) shows a series of magnetic susceptibility for SSs over the entire range, where large Curie-like increases are observed at low temperatures, as more clearly evidenced by the linear temperature dependence in the inverse susceptibilities below 100 K shown in the inset. They come from nearly free spins because the Weiss temperature is almost zero between -0.5 and 0 K. As the Ti content increases, the Curie term first increases, saturates at 30–60%, and then decreases to vanish at $TiO_2$. The *x* dependence of the density of free spins $n_f$ per metal deduced by assuming spin 1/2 with $g = 2$ is plotted in Fig. 7(b). The $n_f$ shows a dome shape with the initial curve near $x = 0$ proportional to *x*, which is simply expected as one orphan spin is generated when a $V^{4+}$–$V^{4+}$ 'molecule' is transformed into a $Ti^{4+}$–$V^{4+}$ 'molecule'.

Following the analyses by Hörlin and coworkers [51], two models are considered: model A assumes a strong lattice dimerization and model B does uniform $V^{4+}$ chains of various lengths terminated by $Ti^{4+}$ at the ends. The former gives $n_f = x(1 - x)$ by taking into account the possibility of creating a $Ti^{4+}$–$Ti^{4+}$ 'molecule' and the latter gives $n_f = x(1 - x)/(2 - x)$ by summing up contributions from all chains with odd number of V atoms which may be assumed to carry one nearly non-interacting spin. Our $n_f$ data follows model A at $0 < x < 0.2$, which is consistent with the presence of the transition to the M1 phase. Then it deviates from model A in the intermediate *x* range and approaches model B only near $x = 1$, suggesting that there remains a short-range correlation toward dimerization in the intermediate *x* range in spite of the absence of global structure symmetry breaking. Presumably, a dimerization takes place to form a metal–metal bonding when two V atoms meet each other, which distorts the surrounding only locally to induce $Ti^{4+}$–$V^{4+}$ 'molecules' carrying orphan spins. Therefore, all the impurity effects on the insulating phase are well understood within Goodenough's molecular–orbital picture, illustrating the primary importance of the structural dimerization of the local character for the mechanism of the MIT of $VO_2$.

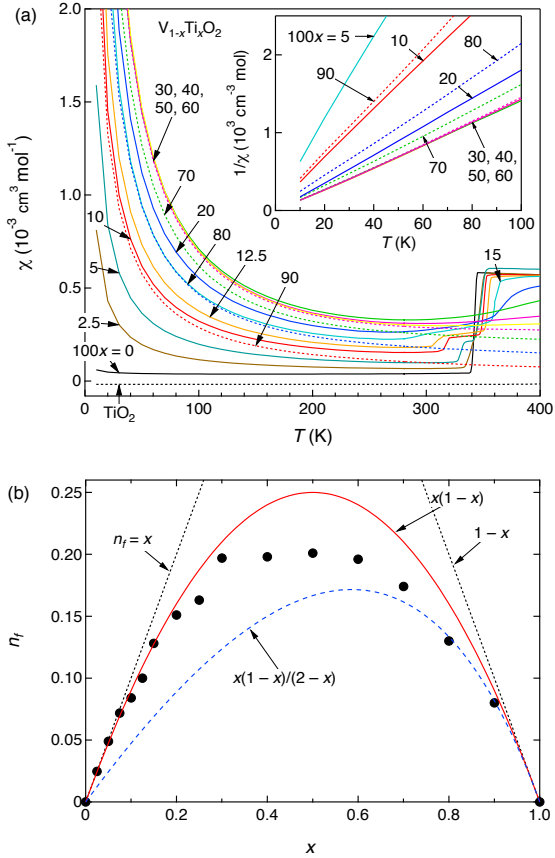

Fig. 7. (a) Magnetic susceptibilities measured at magnetic field of 1 T upon cooling below 400 K for solid solutions $V_{1-x}Ti_xO_2$ in the main panel and their inverse below 100 K in the inset. The data for $x \leq 0.5$ and $> 0.5$ are represented by solid and dotted lines, respectively. (b) Number of free spins per metal $n_f$ as a function of the Ti content $x$. The solid, broken and dotted lines represent calculated $x$ dependences. See text for detail.

## 2.2. Rutile-related compounds

Let's look at $VO_2$ from a broad perspective of crystal chemistry for the family of rutile and its related compounds. Transition metal dioxides that crystallize in the rutile or rutile-related structures are listed in Table 1. Early transition metal elements are favored except for Zr and Hf in the $d^0$ electron configuration. Fe, Co, and Ni in the $3d$ row are also excluded, because they prefer the trivalent or divalent state in different crystal structures. Cr, Mn, Ru, Os, and Ir dioxides are stable in the rutile type [67], while Rh, Ta, and Pt dioxides seem less stable. The preparations of $RhO_2$ and $PtO_2$ require high-pressure conditions [68]. $PtO_2$ may exist in three modifications and the β' form prepared under high pressure adapts the rutile structure [69]. $TaO_2$ may take the rutile structure [70], though there was one report based on electron diffraction experiments to claim that it takes a $NbO_2$-type distorted structure [71].

**Table 1**
Part of the periodic table showing the occurrence of transition metal dioxides in the rutile-type structures. Those except for Fe, Co, Ni, Zr, Pd, and Hf crystallize in the rutile-type structures. Among them, V and Nb dioxides crystallize in distorted structures below the MIT temperatures, and Mo, Tc, W and Re dioxides do at all temperatures. Their structural parameters are listed in Tables 2 and 3.

| Electron Config. | $d^0$ | $d^1$ | $d^2$ | $d^3$ | $d^4$ | $d^5$ | $d^6$ |
|---|---|---|---|---|---|---|---|
| $3d$ | Ti | V | Cr | Mn | Fe | Co | Ni |
| $4d$ | Zr | Nb | Mo | Tc | Ru | Rh | Pd |
| $5d$ | Hf | Ta | W | Re | Os | Ir | Pt |

Among the other compounds, $VO_2$ (below $T_{MI}$ = 340 K), $MoO_2$ [72], $TcO_2$ [73], and $WO_2$ [72] crystallize in a monoclinic structure with space group $P2_1/c$, which is called the $MoO_2$ type or the M1 structure for $VO_2$ [17,67]. The monoclinic unit cell has lattice parameters of $a_m = 2c_r$, $b_m = a_r$, $c_m = a_r/\sin\beta$, $\beta \sim 120°$, where $a_r$ and $c_r$ refer to the rutile-type subcell. A striking feature of this monoclinic phase is the presence of metal–metal pairs in the strands of edge-sharing octahedra along the $c_r$ axis: alternations are large, eg. for $VO_2$ 2.65 and 3.12 Å at 298 K below the MIT in place of the regular 2.87 Å spacing in the tetragonal structure at 360 K [67,74]. $NbO_2$ in its insulating state below $T_{MI}$ = 1070 K takes a more complex tetragonal structure containing 32 molecules per unit cell with space group $I4_1/a$; $a_t = 2\sqrt{2}a_r$ and $c_t = 2c_r$ [75]. Nevertheless, it includes similar metal–metal pairs along the $c_r$ axis as found in the monoclinic structure. α-$ReO_2$ seems to be a metastable phase with metal–metal pairs [67]. Thus, these six compounds crystallize in distorted structures that are commonly characterized by the occurrence of metal–metal pairs.

The structural parameters of the rutile and distorted compounds are listed in Tables 2 and 3, respectively. In Table 2 all known metal dioxides that contain main group elements and crystallize in the rutile structure are also listed for comparison. In the rutile family there is another class of compounds AA'$O_4$ with two kinds of metals randomly distributed [76,77], whose structural parameters are listed in Table 4. Trirutile oxides such as $CuSb_2O_6$ [78] with the three-fold $c$ axis due to the ordering of two kinds of metal atoms are not included in the present paper, because they may have a different crystal chemistry.

Figure 8 shows the cell volume $V$ per formula unit as a function of the effective ionic radius $r_i$ for the six coordination [79]. All the compounds except for $TeO_2$ fall on a universal curve of $V = 41.9(7) + 89(3)r_i^3$; a similar relation was already noted by Shannon [68]. Impressively, the rutile structure is so flexible as to accommodate such a wide range of cations from the smallest $Si^{4+}$ ion in stishovite, the high-pressure form of $SiO_2$, to the large $Pb^{4+}$ ion; both $r_i$ and $V$ vary almost by a factor of two. This is quite exceptional among many structure types for inorganic compounds and is closely related to the specific crystal structure, which will be discussed later. Note also that the compounds with the distorted structures fall on the universal curve, suggesting that the distortions occur so as to keep the cell volume. $TeO_2$, which has the largest volume (87.009 Å$^3$) in the family of rutile oxides, is not on the curve with a given $r_i$ for $Te^{4+}$ ion (0.97 Å) (it is not plotted in Fig. 8); the universal curve would predict $r_i = 0.80$ Å for $Te^{4+}$. This exception is probably due to the $5s^2$ lone-pair electron on $Te^{4+}$.

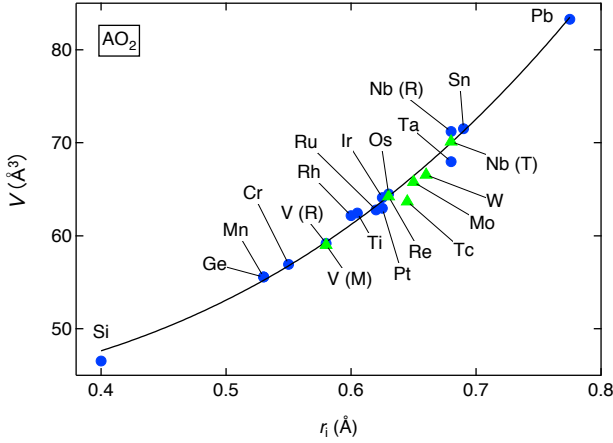

Fig. 8. Cell volume $V$ per formula unit for $AO_2$ compounds plotted against the effective ionic radius $r_i$ for the octahedral coordination. Those with undistorted and distorted rutile structures are shown by circles and triangles, respectively. The solid curve is a fit to the cubic form: $V = 41.9(7) + 89(3)r_i^3$.

All the rutile-related compounds are shown in the $a$–$c$ space of Fig. 9, where $a$ and $c$ are the lattice constants of the tetragonal rutile structure, or those of the distorted structures approximated by considering the relations between the unit cells: $a = (b_m + c_m \sin\beta)/2$ and $c = a_m/2$ for the monoclinic $MoO_2$ type [67] and $a = a_t/(2\sqrt{2})$ and $c = c_t/2$ for the tetragonal structure of $NbO_2$ [75]. Some rutile-type fluorides are also shown for comparison with lattice constants from the literature [77]. Apparently, the rutile compounds exist in the wide ranges of both $a$ and $c$: the $c$ ($a$) value can vary in a wide range at a similar $a$ ($c$) value, demonstrating an enormously flexible nature of the rutile structure. For example, it is intuitive to compare two rutile compounds, $VO_2$ and $RuO_2$. Although the ionic radii are larger for $Ru^{4+}$ (0.62 Å) than $V^{4+}$ (0.58 Å), the $a$ is smaller in $RuO_2$ than in $VO_2$, and instead, the $c$ is much larger in $RuO_2$. Thus, the difference in the ionic radii of the metal ions is absorbed by expanding the space along the $c$ axis, which exemplifies an interesting feature of the rutile structure, as will be mentioned in detail in the next section. On the other hand, Re and Os, which are neighbors in the periodic table and thus take similar ionic radii, are located far apart in Fig. 9, though their dioxides have similar cell volumes (Fig. 8). This indicates that the ionic radius is not a critical parameter for structural stability.

It is important to note in Fig. 9 that the undistorted rutile compounds lie at the upper-left space, while the distorted compounds lie at the lower-right space. Very interestingly, two MIT compounds, $VO_2$ and $NbO_2$, in the rutile form are located in between them: they happen to have nearly the same axial ratio of $c/a = 0.625$, and the borderline of $c/a =$ 0.625 seems to separate the two structural groups. Moreover, the $c/a$ ratios of the monoclinic $VO_2$ and the tetragonal $NbO_2$ are slightly larger and smaller than those of the corresponding rutile phases, respectively, but their differences are relatively small; the translations of the lattice constants of the distorted structures into the rutile one are approximate and may cause some deviations. Thus, the two structures must be nearly degenerate at the two compounds near the borderline. This observation makes one believe that there is an inherent structural instability for the two compounds that causes a transition between the two structures. The importance of the axial ratio on the crystal chemistry of rutile-type compounds has already been pointed out by many solid state chemists [14,17,67,80,81] and will be further discussed in the present paper.

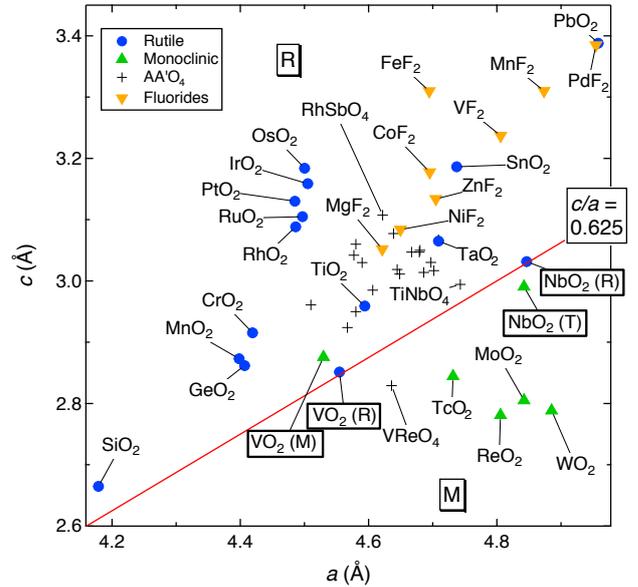

Fig. 9. Rutile-related compounds of $AO_2$ (circle and triangle), $AA'O_4$ (cross) and $AF_2$ (upside-down triangle) plotted in the $a$–$c$ space. The borderline approximately dividing the rutile (R) and distorted structures (M, mostly monoclinic) possesses the axial ratio of $c/a = 0.625$, on which two MIT compounds $VO_2$ and $NbO_2$ are located.

It is further noted in Fig. 9 that another $d^1$ compound $TaO_2$ exists apart from the borderline in the rutile side. The properties of $TaO_2$ are not clear because of difficulty in sample preparation [82] but can be metallic according to band structure calculations [83]. In addition, there is one exceptional compound $VReO_4$ [84], which was prepared under high pressure of 6 GPa, with an undistorted rutile structure in spite of the fact that it is located below the borderline. Possibly, the random distribution of V and Re atoms stabilizes the undistorted structure, as metal–metal pairing tends to be disturbed.

Transition metal dioxides with the rutile-type structures possess a wide variety of electrical and magnetic properties [21,85,86]. This is particularly striking among the $3d$ compounds; a wide-gap semiconductor ($TiO_2$), an antiferromagnetic semiconductor or semimetal ($MnO_2$), and a metallic ferromagnet ($CrO_2$). By comparison, most $4d$ and $5d$ compounds are found to be metallic as in $RuO_2$ except for NbO$_2$ at low temperatures below $T_{MI}$. These chemical trends have been interpreted as a function of $d$-electron number [14,21,23,67]. It is emphasized here that the distorted structures appear irrespective of electronic properties: VO$_2$ and NbO$_2$ with the $d^1$ configuration happen to become insulators in the distorted structures, while MoO$_2$, TcO$_2$, WO$_2$ and α-ReO$_2$ in the same or similar structures remain metals down to the lowest temperature [67].

### 2.3. Crystal chemistry

#### 2.3.1. Rutile structure

The rutile structure is geometrically simple [87]. It crystallizes in the tetragonal structure of space group $P4_2/mnm$ with the cations at the positions of (0, 0, 0) and (1/2, 1/2, 1/2) and the four anions at ±($x$, $x$, 0) and ±(1/2 + $x$, 1/2 − $x$, 1/2) in the tetragonal unit cell. Thus, the geometry is determined by only two parameters, $c/a$ and $x$. In the structure linear strands of edge-sharing AX$_6$ octahedra along the $c$ axis are present at the corner and the center of the basal square; the two strands are joined with each other via octahedron vertices after the halfway shift along the $c$ axis so as to align in a staggered way (Fig. 1). Since the metal site takes the local symmetry of $mmm$, only mirror planes without rotation axes, the octahedron is not necessarily regular but always orthorhombically distorted, depending on the $c/a$ ratio and the $x$ coordinate; there are four equatorial O atoms that form a rectangle and two apical O atoms with the A–O bond perpendicular to the $c$ axis. This is the reason why the rutile structure can accommodate such a wide range of cations without breaking the crystal symmetry. The AO$_6$ octahedron would become regular, only when $c/a = 2 - \sqrt{2} \sim 0.5858$ and $x = (c/a)/2 \sim 0.2929$, which is never achieved in the actual compounds of the rutile type as $x > 0.29$ and $c/a > 0.60$.

In order to shed light on the characteristics of the rutile structure, two typical compounds, VO$_2$ and RuO$_2$, are visually compared in Fig. 10, which are schematically depicted with the effective ionic radii: $r_i$ = 0.58, 0.62 and 1.40 Å for V$^{4+}$, Ru$^{4+}$ and O$^{2-}$ ions [79]. As mentioned above, the $a$ is shorter and the $c$ is longer for RuO$_2$, and the increase in $r_i$ is adjusted by elongating the $c$ axis length. As depicted in Fig. 10, the V$^{4+}$ and Ru$^{4+}$ ions fit well the voids of the rectangle of oxide ions. On the other hand, neighboring oxide ions are far separated along the $c$ axis, while compressed to overlap with each other along the [110] direction. Such a significant overlap between oxides ions is rarely observed in other metal oxides. This evidences a strong covalent bonding in the pair of oxide ions in the [110] direction, which is more enhanced from VO$_2$ to RuO$_2$. Thus, chemical bonds in the rutile oxides are not purely ionic and take a certain covalent character because of the high valence state of the A$^{4+}$ ions. It is noted, however, that covalency occurs mainly at the bridging O ions, which prevents a direct metal–metal bonding along the $c$ axis. The presence of the covalent pair of oxide ions suggests a tendency to form a peroxide ion O$_2^{2-}$ as in BaO$_2$ with some charge transfer to the metal ions.

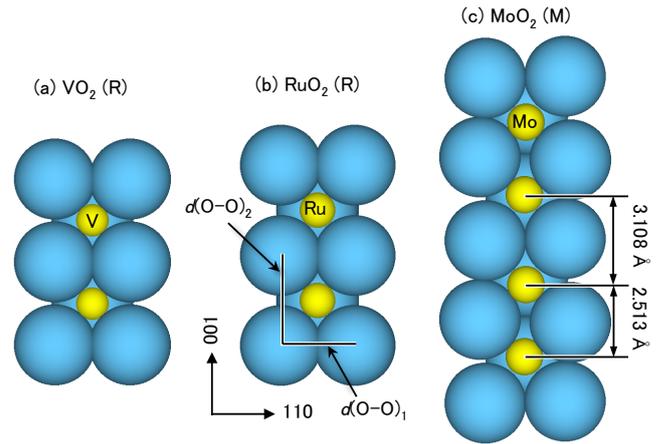

Fig. 10. Crystal structures depicted by spheres with effective ionic radii (0.58, 0.62, 0.65 and 1.40 Å for V$^{4+}$, Ru$^{4+}$, Mo$^{4+}$ and O$^{2-}$ ions, respectively) for (a) VO$_2$ and (b) RuO$_2$ both in the rutile form and (c) monoclinic MoO$_2$. They represent cross-sectional views of the strands of octahedra at the center of the tetragonal unit cell along the [1–10] direction (Fig. 1). A corresponding projection is shown for MoO$_2$ with a monoclinic unit cell.

In the case of more ionic anions like the chloride ion, such compression along the [110] direction becomes difficult, so that the strands of edge-sharing octahedra tilt with each other, resulting in the orthorhombic CaCl$_2$-type structure; the space group of the CaCl$_2$ type is $Pnnm$, a subgroup of $P4_2/mnm$. On the other hand, small fluoride ions can adapt the short distance within the ionic bonding character and is well stabilized in the tetragonal structure. In contrast, the oxide ion with an intermediate size has to be squeezed to fit in the rutile structure. SiO$_2$ crystallizes in the rutile structure under high pressure but is transformed to a CaCl$_2$-type structure when further compressed above 100 GPa [88], as no more compression of the O pairs may be possible. Moreover, PtO$_2$ seems to take either the rutile or the CaCl$_2$-type structure when it was prepared under high pressures [68,69]. Thus, the rutile-type oxides must experience certain structural instabilities for some metal elements. The introduction of the distorted monoclinic or tetragonal structures with metal–metal pairs may be one way to relax the rutile structure.

#### 2.3.2. Structural instability in the rutile compounds

The structure and chemical bonding in the rutile compounds have been discussed by many researchers through the improvement of sample preparation techniques and structural analyses [67,80,81,89,90]. Based on the structural data, for example, Rogers et al. and Baur and Khan made systematic comparisons among the rutile compounds and pointed out the importance of $d$-electron configurations on the crystal chemistry [67,80]. More recently, Bolzan and coworkers obtained reliable structural data on some rutile oxides using the powder neutron diffraction technique and compared them with each other [81].

Thanks to accumulated crystallographic data for rutile-related compounds, I would like to discuss the crystal chemistry in more detail and try to deduce meaningful information about a possible structural instability that is inherent to the rutile structure. Figure 11 plots the $x$ coordinate of anions as a function of $c/a$ for the undistorted rutile compounds; similar plots have already been given [80,81,90]. Most data except for $NbO_2$ were obtained at around room temperature, whereas the $NbO_2$ data was obtained at 1273 K above the $T_{MI}$ [91]. Although there is a positive temperature dependence for $c/a$ (see Fig. 19), it is negligible compared with the range plotted in figure [30]. Moreover, the temperature dependence of $x$ may be also negligible. In fact, changes in $TiO_2$ are quite small; 0.002 for $c/a$ and 0.0006 for $x$ between 300 and 1200 K [92]. Thus, comparisons of the $NbO_2$ data with the others make sense.

It is noticed in Fig. 11 that the $x$ values of the $AO_2$ compounds are rather constant at around 0.305, insensitive to the A element as well as the $c/a$ ratio. Note that $SiO_2$ and $PbO_2$ possess almost same values in spite of the large difference in the ionic radius between $Si^{4+}$ and $Pb^{4+}$ ions. $TaO_2$ and $CrO_2$ have slightly smaller $x$ values than others. Some fluorides and chlorides also take values between 0.301 and 0.306; $CaCl_2$ takes the rutile form at high temperature above 508 K [93]. Interestingly enough, compared with other compounds, $VO_2$ and $NbO_2$ take significantly smaller values of 0.3001 and 0.2924, respectively, as already noted by Baur [90], McWhan et al. [74], Rogers et al. [67], and Bolzan et al. [91], clearly indicating that the two compounds already suffer a certain instability on the verge of the stability range of the rutile type [35,74,91]. In addition, two $AA'O_4$ compounds with small $c/a$ values also have small $x$ values: 0.294 for $VReO_4$ with $c/a$ = 0.610 and 0.295 for $TiNbO_4$ with $c/a$ = 0.631.

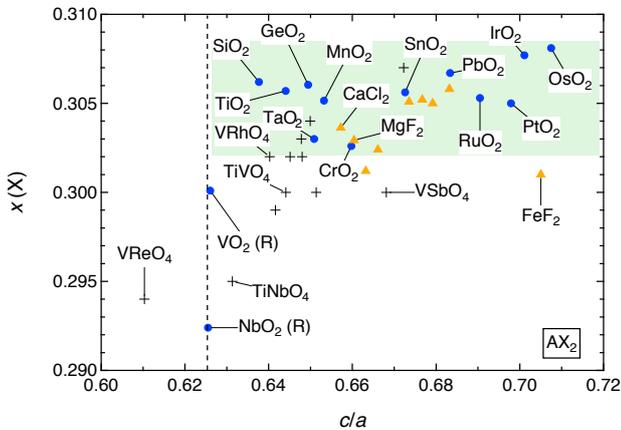

Fig. 11. Coordinate $x$ of the anions for the rutile compounds plotted against the axial ratio $c/a$. $AO_2$ compounds are shown by circles, $AA'O_4$ compounds by crosses, and fluorides ($MgF_2$, $VF_2$, $FeF_2$, $CoF_2$, $NiF_2$, $MnF_2$, $ZnF_2$, $PdF_2$) [77] and chlorides ($CaCl_2$) [93] by triangles.

Chemical trends of some selected bond lengths as a function of $c/a$ for all the $AO_2$ compounds are shown in Fig. 12. First I would like to describe those in the rutile region with $c/a$ > 0.625. Two distances, $d(A–O)_1$ and $d(A–O)_2$, plotted in Fig. 12(a) are the distances from the A atom to the apical and the equatorial O atoms, respectively, which are visually referred in the inset of Fig. 13(a). $d(A–O)_1$ is shorter than $d(A–O)_2$ at large $c/a$ and vice versa at small $c/a$. Their ratio is plotted in Fig. 13(a), where a systematic increase in the ratio $d(A–O)_1/d(A–O)_2$ with decreasing $c/a$ is clearly observed; a crossover from a compressed octahedron to an elongated one occurs at $c/a \sim 0.66$ with decreasing $c/a$. Some $AA'O_4$ compounds show deviations from the trend for the $AO_2$ compounds, but this may be because of the random distribution of two kinds of cations and also because of less reliable $x$ coordinate values for $AA'O_4$. Note that the distortion of the octahedron is not governed by the Jahn–Teller effect (the $d$-electron number) but by the geometrical constraint of the rutile structure.

Three O–O distances plotted in Fig. 12(b) are $d(O–O)_1$ and $d(O–O)_2$, which are the lengths of the basal O rectangle perpendicular and parallel to the $c$ axis, respectively, and $d(O–O)_3$ between the basal and apical O atoms, as shown in Fig. 13(b). The chemical trends of these O–O distances are rather complex. First look at $d(O–O)_3$, which has nearly constant values at 2.8 Å at large $c/a$ and scattered around the value with decreasing $c/a$. This value is exactly equal to twice the ionic radius of the $O^{2-}$ ion, indicating the ionic character of the bond between the basal and apical O atoms. Compared with this, $d(O–O)_1$ is always much smaller than 2.8 Å, and $d(O–O)_2$ is larger at the large $c/a$ side and they tend to merge at 2.8 Å with decreasing $c/a$. This tendency is what has been illustrated in Fig. 10 for rutile $VO_2$ and $RuO_2$: O ions are squeezed along the [110] direction and far separated along the $c$ axis for $RuO_2$, and the difference is reduced toward $VO_2$. In contrast to the case of dioxides, the $d(F–F)_1$ of difluorides is always close to twice the ionic radius of the $F^-$ ion; no squeezing for the F ions along the [110] direction, indicative of a purely ionic chemical bonding. The rectangular ratio of the basal O atoms for metal dioxides, $d(O–O)_2/d(O–O)_1$, is plotted in Fig. 13(b). In spite of a large scatter of the actual bond lengths, the ratio shows liner behavior against $c/a$. This is not surprising because $d(O–O)_2$ equals $c$, so that one would expect a linear relation when $x$ is constant.

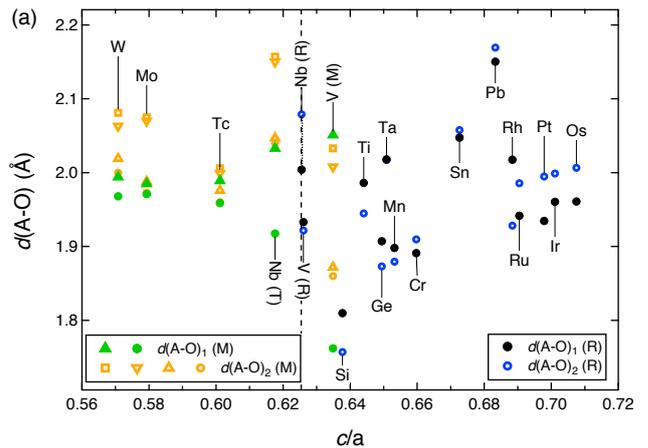

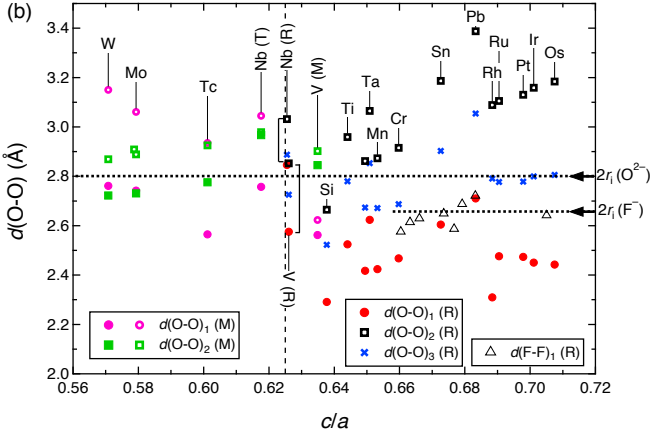

Fig. 12. A–O distances (a) and O–O distances (b) for $AO_2$ compounds with the regular and distorted structures as a function of $c/a$. The F–F distance $d(F-F)_1$ of some metal difluorides that corresponds to $d(O-O)_1$ are also plotted in (b).

Let me summarize the observed trends on the shape of the $AO_6$ octahedron as a function of $c/a$ for the undistorted rutile compounds; although they are natural consequences for nearly constant $x$ values, they are helpful for understanding the origin of the structural instability. As mentioned in the preceding paragraph, the octahedron changes its shape from compressed to elongated with decreasing $c/a$ [Fig. 13(a)]; all the A–O bond lengths happen to be equal at $c/a \sim 0.66$, but the basal O rectangle is elongated by about 20% along the $c$ axis [Fig. 13(b)]. The $d(A-O)_1/d(A-O)_2$ ratio increases steeply toward the critical $c/a$ value of 0.625, which mostly comes from the large decrease of $d(A-O)_2$ with decreasing the $r_i$ of the A ion [Fig. 12(a)]. This means that the octahedral coordination with the $mmm$ symmetry at the A site becomes unstable when the $c/a$ approaches the critical value. Markedly, $VO_2$ and $NbO_2$ near the border take much smaller values than expected from the general trend. In other words, if they followed the relation, their octahedra would experience huge deformations! On the other hand, as shown in Fig. 13(b), the rectangular ratio $d(O-O)_2/d(O-O)_1$ decreases linearly with decreasing $c/a$ from a large value of 1.3 for $OsO_2$ to a small value of 1.16 for $SiO_2$, which mostly comes from the decrease in $c$. Probably, more important is the insensitivity of the $d(O-O)_1$, which remains always short in contrast to the large variation in $d(O-O)_2$. This means that the covalent O–O bonding is always kept in the rutile structure, or it is the requirement for the structure. What happens in $VO_2$ and $NbO_2$ at the critical $c/a$ is a sudden increase in $d(O-O)_1$ [Fig. 12(b)], indicating a tendency to lose the covalent O–O bonding. Therefore, the rutile structure is going to become unstable with decreasing $c/a$ and, at the critical $c/a$ ratio, the stability must be influenced even by a small perturbation.

The presence of a structural instability in $VO_2$ and $NbO_2$ has been inferred by the observation of exceptionally large atomic displacement parameters both in the metal and oxide atoms [74,91]. Moreover, symmetry considerations and the lattice dynamics calculations show that there are soft modes in $VO_2$ and $NbO_2$ at the R point of the Brillouin zone that may trigger structural transitions [94,95]. In fact, a profound softening was observed in the Raman spectrum of rutile $VO_2$ [96] and in the neutron scattering experiments on $NbO_2$ [97]. Thus, it is likely that the soft mode at the R point is frozen in at the transitions [22].

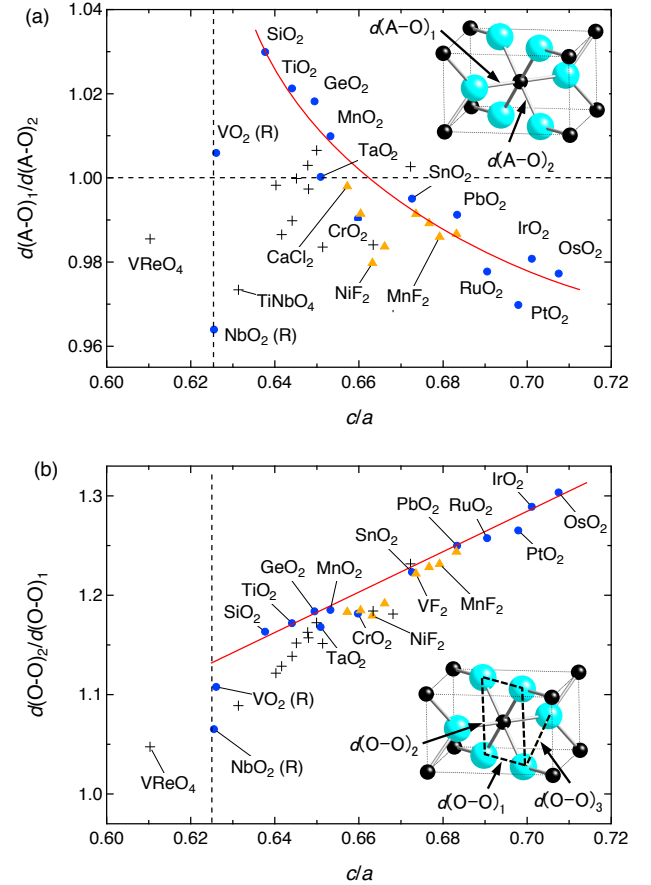

Fig. 13. Two ratios defining the shape of the $AX_6$ octahedron: (a) $d(A-O)_1/d(A-O)_2$ and (b) $d(O-O)_2/d(O-O)_1$. The solid lines are guides to the eye. $AO_2$ compounds are shown by circles, $AA'O_4$ compounds by crosses, and fluorides and chlorides by triangles.

### 2.3.3. Distorted rutile structures

Among the other rutile-related compounds, $VO_2$ (below $T_{MI}$), $MoO_2$, $TcO_2$ and $WO_2$ crystallize in the $MoO_2$ type structure [17]. Figure 10(c) shows a cross-sectional view of $MoO_2$ in a similar way as in the rutile structure. It is clearly observed that the Mo–Mo distance heavily alternates between 2.513 and 3.108 Å (21% difference!) along the rutile $c$ axis. The closer Mo–Mo distance is made possible by pushing the two bridging O ions away from each other; there is no doubt about that a strong metal–metal bond is generated through this open channel. Note that even in the other bridging O pairs the O–O distance is relaxed to have no overlapping between them: the O–O distances are 2.742 and 3.061 Å, both much larger than 2.476 Å for $RuO_2$. Thus, in this monoclinic structure, the covalent character of the bridging O ions has been completely removed. The breaking of the O–O pairs and the generation of a metal–metal bonding should elongate the rutile $a$ axis and compress the $c$ axis, respectively, so that the $c/a$ ratio is reduced; the stronger the metal–metal bonding, the smaller the $c/a$ ratio. Plausibly, the energy gain due to the formation of metal–metal bonds is large enough to overcome the resulting increase in lattice energy.

NbO$_2$ in its insulating state takes a more complex tetragonal structure, which includes two cation sites instead of one site in the monoclinic structure [75,81]; the two Nb sites appear alternately along the $c$ axis. Nevertheless, the deformation pattern in one strand of edge-sharing octahedra resembles that of the MoO$_2$ type, which includes similar metal–metal pairs and shifts of surrounding O ions. α-ReO$_2$ crystallizes in another monoclinic structure and also contains two cation sites as in NbO$_2$, one of which splits into two equivalent sites randomly occupied [67]. Anyway, it is also considered to be a member of the MoO$_2$ type, because there are two kinds of the Re–Re distance of 2.622 and 2.993 Å in the chain.

The A–O and selected O–O distances of the distorted compounds except for α-ReO$_2$ are plotted in Fig. 12. Among six different A–O bonds the two apical bonds are shorter than the four basal bonds (the octahedron is always compressed), as observed for rutile compounds with heavy metals at large $c/a$ ratios. As for the O–O distances of the basal rectangle, $d$(O–O)$_1$ perpendicular to the rutile $c$ axis is much larger in the distorted structures than in the rutile structure, as noted above for MoO$_2$. The shortest $d$(O–O)$_1$ takes values close to $2r_i$(O$^{2-}$) except for the cases of VO$_2$ and TcO$_2$; the two $d$(O–O)$_1$ values of monoclinic VO$_2$ remain nearly equal to that of rutile VO$_2$. On the other hand, the $d$(O–O)$_1$ of rutile NbO$_2$ is already close to $2r_i$(O$^{2-}$) and do not change in the distorted structure. In other words, rutile VO$_2$ and NbO$_2$ are already subject to the distortions toward the dimer structures, reflecting that they are located near the border between the two structural groups. In contrast, the large atomic shifts in MoO$_2$ and WO$_2$ imply that the distorted structure is really stable for Mo and W at reduced $c/a$.

Figure 14 shows the metal–metal distances for all the metal dioxides. A large dimerization takes place commonly in every distorted compound: 16, 20, 16, 13, 21 and 22% for VO$_2$, NbO$_2$, TcO$_2$, α-ReO$_2$, MoO$_2$ and WO$_2$, respectively. The shorter metal–metal distance decreases with decreasing $c/a$, which seems to follow a line drawn for undistorted transition metal dioxides. In contrast, the longer distance remains nearly constant at 3.1 Å, which is close to the value of RuO$_2$ and thus is a natural distance for heavy elements in the regular structure. This means that the reduction of $c/a$ in the distorted structures is mostly accomplished by shortening the shorter bond with the longer bond intact. This evidences that an additional strong chemical bond occurs between the paired cations, which has been ascribed to a direct metal–metal bonding [14,15,35,67].

Marinder and Magnéli first noted the presence of a metal–metal bonding in the MoO$_2$-type compounds [14]. Based on Cotton's empirical correlations of the metal–metal distance vs. the bond order [98], Rogers et al. and Sleight et al. obtained that the bond order is 1 for the short bonds of VO$_2$ and NbO$_2$ and 2 for those of MoO$_2$, TcO$_2$, WO$_2$ and α-ReO$_2$, while 0 for TiO$_2$, CrO$_2$, RuO$_2$ and SnO$_2$ [67,99]. This implies that additional weak metal–metal bonds of the covalent character occur in VO$_2$ and NbO$_2$ with the $d^1$ configuration, while strong ones in MoO$_2$ and others with the $d^2$ or $d^3$ configuration.

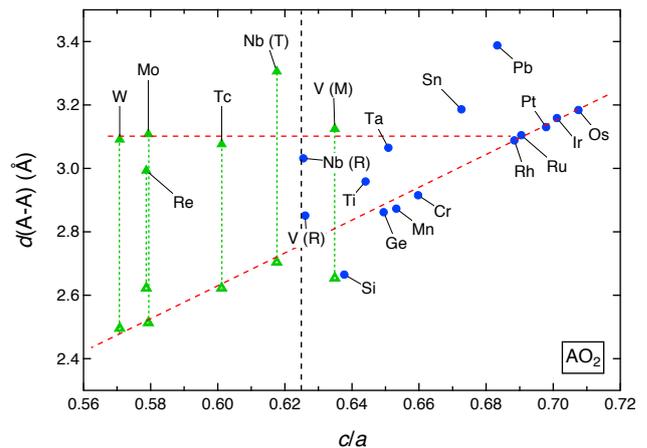

Fig. 14. Metal–metal distances for the rutile (circle) and the distorted structures (triangle) against $c/a$. The broken lines are guides to the eye.

### 2.3.4. M1 and M2 structures of VO$_2$

Until now I have considered only one strand of octahedra, but whether such a dimerization can actually occur or not must depend on the three-dimensional network. Here I consider the M1 and M2 structures of VO$_2$ in more detail. As schematically depicted for the M1 structure in Fig. 15, when a cooperative dimerization takes place in chain A at the origin of the rutile unit cell, chain A' at the center of the unit cell is affected accordingly: as the bridging O atoms between paired V atoms in chain A are pushed outside, the connected octahedron in chain A' is compressed along the [110]$_r$ direction. Same dimerization occurs in the next chain A" at the opposite corner of the unit cell, but there is a halfway shift along the $c_r$ axis between the dimerization patterns of chains A and A". As the result, only one of two apical O atoms in an octahedron in chain A' comes closer to the V atom, which pushes the V atom from the center of octahedron in a staggered way along the $c_r$ axis. Moreover, since the identical dimerization occurs in chain A' too, the V atom moves along a slightly inclined direction from the [110]$_r$ to the $c_r$ axis, as depicted in Fig. 15. Therefore, in order to balance all the atomic movements, the dimerization of V atoms turns out to be accompanied by an antiferroelectric shift. This antiferroelectric distortion must be important for the MIT of VO$_2$, because it raises the π* band above the Fermi level [18,21]. However, note that the lateral displacement is just a consequence of the pairing along the $c$ axis.

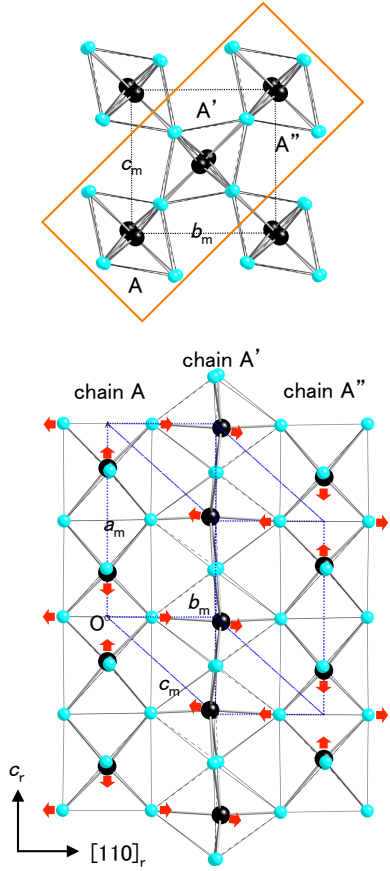

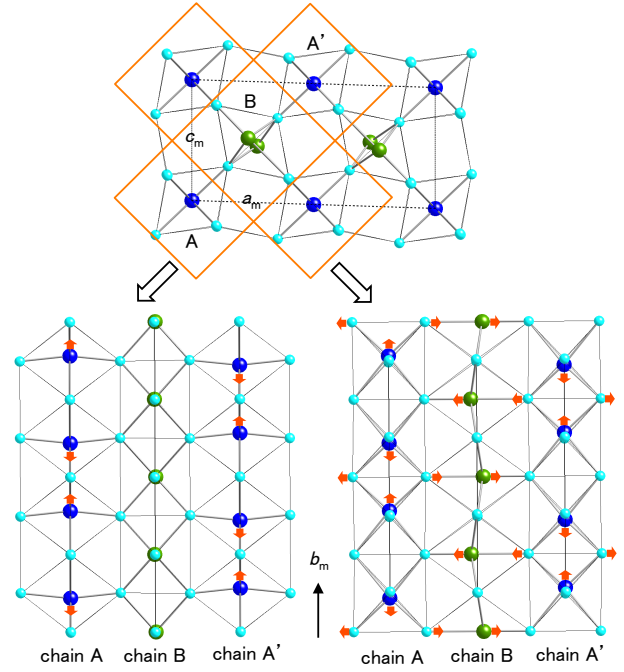

Fig. 15. Schematic representations of the M1 structure of VO$_2$. A structure viewed along the $a_m$ axis of the monoclinic unit cell which corresponds to the $c_r$ axis of the rutile cell is shown at the top and a cross-sectional view of the structure in the rectangle of the top drawing is shown at the bottom. The directions of atomic shifts from the original positions of the rutile structure are approximately indicated by arrows.

In the M2 structure shown in Fig. 16, on the other hand, a cooperative dimerization takes place only in one set of chains; the M2 structure appears with the substitutions of Cr, Ti and others for V or when an uniaxial pressure is applied along the [110]$_r$ direction [100]. Since it is stabilized by minimal substitutions such as 0.3% or under a small uniaxial pressure, it must be a metastable structure with nearly same free energy as the rutile and the M1 phases, and is assumed to be a compromise structure between them. As the result of the dimerization in chain A, an antiferroelectric shift of V atoms occurs in chain B, as in the M1 structure. In contrast to the M1 structure, however, chain B becomes a zigzag chain with a uniform V–V distance without dimerization. Note that such an antiferroelectric shift should not occur in chain A because of the absence of dimerization in chain B. Thus, in both M1 and M2 structures, V–V pairing is made possible by absorbing the large lateral movements of the bridging O atoms in smart ways to minimize the increase in lattice energy in the three-dimensional networks. Obviously, the chain-like structure of the parent rutile structure is advantageous for this.

Fig. 16. Schematic representations of the M2 structure of VO$_2$. A structure viewed along the $b_m$ axis of the monoclinic unit cell which corresponds to the $c$ axis of the rutile cell is shown at the top and cross-sectional views of the structures in the rectangles of the top drawing are shown at the bottom. The directions of atomic shifts from the original positions of the rutile structure are approximately indicated by arrows.

## 2.4. Related compounds

### 2.4.1. Metal–metal bonding in transition metal compounds

Metal–metal bonding is not specific to the MoO$_2$ type structure but occurs in some transition metal compounds containing strands of edge- or face-sharing octahedra [101]. Although high symmetry is generally preferred by the point charge model, a crystal often distorts to have a low-symmetry structure in order to gain more chemical bonding energy. There are two major reasons for this: one is the nonlinearity of the chemical bonding energy against distance [102] and the other is a stabilization by forming an additional direct metal–metal bonding [14,15]. In the former mechanism a cation can shift from the center of octahedron to establish stronger chemical bonds with fewer ligands at shorter bond lengths, which results in a distortion of octahedron, as found in BaTiO$_3$ [102]. This effect may be important for a strongly covalent bond in the $d^0$ configuration. On the other hand, when a shift of a cation in a polyhedron occurs so as to form a pair with that in the neighbor, a further stabilization is expected by generating a metal–metal bond. The formation of metal–metal pairs reminds us of a homopolar chemical bond in a hydrogen molecule, where a bonding singlet state arises as the ground state with an antibonding triplet state as the excited state in the classical Heitler–London scheme.

When metal–metal bonds form in pairs between metal atoms of adjacent octahedra, the metal atoms shift from the octahedron centers in the direction of the corresponding octahedral face or edge, and the octahedron experiences a large distortion [101]. Typical examples are found in the face-sharing strands of octahedra of β-TiCl$_3$ ($3d^1$), ZrI$_3$ ($4d^1$), MoBr$_3$ ($4d^3$), RuCl$_3$ ($4d^5$) or RuBr$_3$ ($4d^5$) and in the edge-sharing strands of octahedra of NbCl$_4$, NbI$_4$ ($4d^1$) or WCl$_4$ ($5d^2$). For example, MoBr$_3$ possesses a face-sharing strand embedded in a hexagonal closest-packing of Br atoms, as shown in Fig. 17(a). Mo–Mo pairs occur in the strand to result in a large alternation in the interatomic distance: 2.875 and 3.203 Å (10.8% difference) [103]. The formation of the Mo–Mo bond causes a large deformation of octahedron with the three bridging Br atoms across the bond moving outward and the other three Br atoms inward.

The crystal structure of NbCl$_4$, shown in Fig. 17(b), gives a more intriguing resemblance to the MoO$_2$ structure. It is monoclinic with space group $I2/m$ and consists of two nearly orthogonal strands of oxygen octahedra when viewed along the $b_m$ axis, which is basically the same as the rutile structure. However, octahedra in one set of the strands are vacant without cations inside of them, so that the Nb–Cl strands are not directly connected to each other. This isolated strand made of NbCl$_6$ octahedra exhibits a similar dimerization as observed in the MoO$_2$ structure as a result of strong metal–metal bonding. The bond alternation is huge; 3.029 and 3.794 Å (22.4% difference) [104]. Moreover, the two bridging Cl atoms at the shorter bond are heavily pushed outward, as obviously noticed in Fig. 17(b). Thus, the structure of the NbCl$_4$ chain resembles those of the MoO$_2$ type, illustrating a general trend forming a metal–metal bonding in the related structures. From another perspective, its atomic displacement pattern would be quite similar to that of the M2 structure of VO$_2$, if there was another Nb atom in the vacant octahedron. The structure of WCl$_4$ is nearly the same as NbCl$_4$, while NbI$_4$ has similar dimerized chains in a different packing from NbCl$_4$ [101]. In contrast, OsCl$_4$ does not show a dimerization even in the nearly same crystal structure as NbCl$_4$; the Os–Os distance is 3.560 Å [105]. This may be because the $d$ orbitals have contracted enough that orbital overlap is poor, and metal–metal bonding no longer occurs [105]. In addition, it may be related to the $5d^4$ electron configuration of the Os$^{4+}$ ion, in which two electrons occupy the lower $d_{xy}$ orbital with two in the upper $d_{yz}/d_{zx}$ orbitals in an octahedron compressed along the local $z$ axis perpendicular to the chain, so that little energy gain is expected by forming a metal–metal bond from the $d_{xy}$ orbitals.

These comparisons between the distorted rutile structures and other different classes of compounds make it clear that the metal–metal bonding is quite ubiquitous in transition metal compounds having chain structures with edge- or face-sharing octahedra. An important issue to be pointed out here is that commonly in these structures the deformations of one strand of octahedra that allow bond alternations are easily accommodated by the surrounding strands in consistent ways so as to minimize the increase in lattice energy. Thus, even a small electronic energy gain by forming a local bonding state in a pair of $d$ orbitals can stabilize the distorted structures.

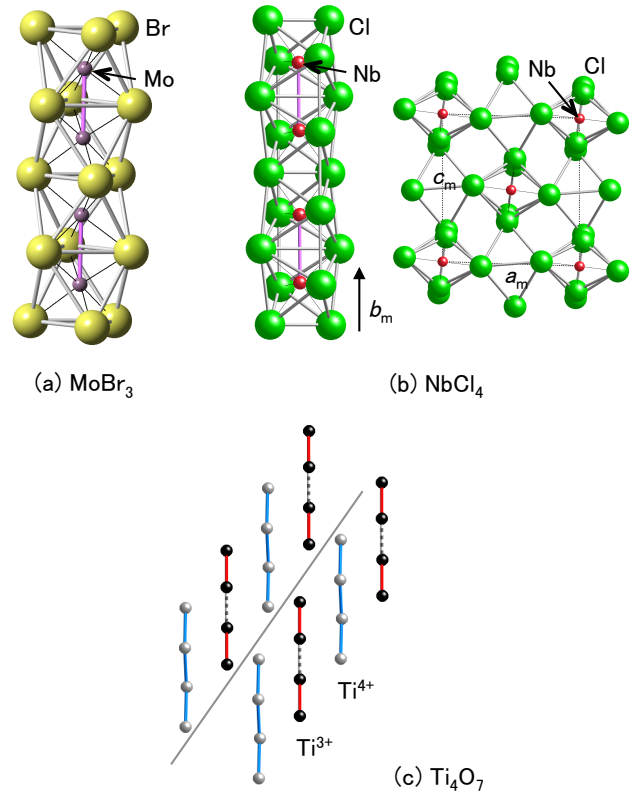

Fig. 17. Crystal structures of related compounds: (a) MoBr$_3$ and (b) NbCl$_4$ at room temperature, and (c) Ti$_4$O$_7$ at 120 K in its insulating phase. In the strand of face-sharing octahedra of MoBr$_3$, Mo ions occur in pairs at a distance of 2.875 Å with 3.203 Å apart [103]. The structure of NbCl$_4$ in (b) resembles the rutile structure when viewed along the monoclinic $b_m$ axis (right), and in the edge-sharing octahedral strand a Nb chain is heavily dimerized with distances of 3.029 and 3.794 Å (left) [104]. In the ultrathin rutile slab of the Magnéli phase Ti$_4$O$_7$ of (c), Ti$^{3+}$ chains are dimerized with distances of 2.802 and 3.133 Å, while Ti$^{4+}$ chains remain nearly uniform in between the Ti$^{3+}$ chains [37].

### 2.4.2. Related MIT compounds: Magnéli phases and $V_2O_3$

Ti$_4$O$_7$ is one of the Magnéli phases containing equal proportion of Ti$^{3+}$ and Ti$^{4+}$ cations [34]. It crystallizes in a rutile-related structure consisting of strands of edge-sharing octahedra which are truncated every four octahedra by crystallographic shear planes, as schematically depicted in Fig. 17(c) [37]. The compound shows an MIT at 149 K, followed by a second transition at 125 K that involves some structural rearrangement. Above 149 K all the $d$ electrons are delocalized over Ti$^{3.5+}$, while below 125 K they are completely localized into alternate chains of Ti$^{3+}$ sites separated by Ti$^{4+}$ chains. Dimerization occurs only in the Ti$^{3+}$ chains into non-magnetic singlets with a bond alternation of 2.802 and 3.133 Å (11.2%), while keeping the other Ti$^{4+}$ chains nearly uniform [37]. Note that these changes take place in a pile of ultrathin rutile-like layers of four octahedra-thick and also that the low-temperature structure resembles the M2 structure of VO$_2$, even though the distributions of $d$ electrons are completely different between Ti$_4$O$_7$ and VO$_2$.

A similar situation is observed in V$_4$O$_7$ below $T_{MI}$ = 250

K [40] and in $V_5O_9$ below $T_{MI}$ = 135 K [41]. In $V_4O_7$, $V^{3+}$ and $V^{4+}$ chains are formed below $T_{MI}$ in the same four octahedra-thick layer as in $Ti_4O_7$, both of which exhibit bond alternations of 5.5 and 11.9%, respectively [40]. In $V_5O_9$ with rutile-like layers of five octahedra-thick, a similar charge localization takes place but with more complex atomic displacements [41]. Other Magnéli phases with thicker rutile-like slabs, $Ti_nO_{2n-1}$ ($n$ = 5, 6) [34] and $V_nO_{2n-1}$ ($n$ = 6, 8, 9) [35], also exhibit similar MITs, but their details are not known. It is likely that they can be analogous to those of $Ti_4O_7$ and $V_4O_7$ or closer to that of $VO_2$ for $V_nO_{2n-1}$. These observations on the MITs of the Magnéli phases remind us of a sharp MIT observed in thin lamellas in the spinodally decomposed $(V,Ti)O_2$ sample, suggesting a strong local character in common.

Chakraverty discussed the MITs of $Ti_4O_7$ in terms of bipolaron that is a composite of two electrons of opposite spins coupled together tightly with a strong cloud of local distortion [42]. He pointed out the importance of local lattice deformation due to strong electron–phonon couplings. However, I believe that there must be a common origin for the MITs and structural transitions between $VO_2$ and the Magnéli phases which should not depend on the details of electronic structures such as Fermi-surface geometry or electron filling.

Another nearby, well-known MIT compound is $V_2O_3$, which shows an MIT at 150–160 K between the high-temperature metallic phase with the corundum structure and the low-temperature semiconducting phase with a monoclinic structure [106]. In the corundum structure the vanadium atoms occupy two-thirds of the octahedral holes in an approximately hexagonal closed-packing of oxide atoms. Each $VO_6$ octahedron is connected to another by face sharing and to three more by edge sharing. Thus, there are already built-in pairs of V atoms in the metallic phase. At the MIT on cooling the V–V distance across the shared octahedral face increases from 2.697 to 2.745 Å (+1.8%), and those across the shared octahedral edge increases from 2.882 to 2.987 Å (+3.6%) [107]. Thus, there is no tendency to form a metal–metal bond in $V_2O_3$ below $T_{MI}$, suggesting that the mechanisms of the MITs are substantially different between $VO_2$ and $V_2O_3$. Probably, electron correlations are crucial on the MIT of $V_2O_3$: they give a repulsive force between electrons and must prefer a larger separation for V atoms in the insulating state with less screening [3].

## 2.5. Mechanism of the MIT of $VO_2$

Before starting discussion on the mechanism of the MIT of $VO_2$, let me clarify the meaning of identifying a mechanism. As mentioned in the introduction, the MITs in transition metal compounds are always found out to be complex phenomena as a result of competition among various factors. To specify one of them as a main source would require a quantitative analysis that gives how much free energy is gained by each factor; the one that gets the largest energy gain would win. However, this is almost impossible because they cannot be separated from each other in any way. Thus, it is crucial to carry out a systematic and qualitative analysis over many experimental results and to find one reasonable answer that can explain most of them within a persuasive scenario. Generality of the mechanism to other related phenomena or systems may be also important.

### 2.5.1. Electron correlations or Peierls instability?

The factors governing the MIT of $VO_2$ may be classified into three groups: electron correlations [24,25], Peierls (FS) instability [22,23], or a lattice distortion [18,21,108]. Many review papers are available to know the detail of previous discussion from these points of view; for example, look at the recent one by Eyert [23]. Here I would like to refer them briefly to smoke out a possible suspect.

Aside from the true driving force, the molecular orbital picture by Goodenough has been proved to be qualitatively right in the previous study, the idea of which is shown in Fig. 18 [18,21,45]. The fundamental electronic structure of $VO_2$ is composed of two states originating from the $t_{2g}$ orbitals in the octahedral crystal field: the $d_{//}$ orbitals spreading along the $c$ axis are nearly nonbonding and possess a quasi-one-dimensional character, while the $\pi^*$ orbitals strongly hybridizing with the O $2p$ $\pi$ state are isotropic. Both states are partially occupied in the metallic $VO_2$ at high temperature. In the M1 phase of $VO_2$, the pairing of V atoms causes a bonding–antibonding splitting of the $d_{//}$ band, with the lower $d_{//}$ band fully occupied and the upper $d_{//}$ unoccupied. At the same time, the antiferroelectric shifts of V atoms perpendicular to the $c$ axis, which is naturally caused by the pairing as described for the M1 structure in Fig. 15, result in an upshift of the $\pi^*$ bands above the Fermi level [24,25,109], so that the $\pi^*$ band also becomes completely vacant, opening an energy gap of ~0.7 eV [109,110]. The Goodenough model is intuitive and helpful for understanding what happens at the MIT but it does not tell us about the driving force.

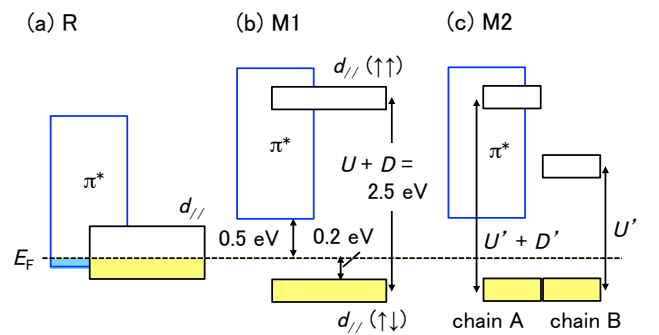

Fig. 18. Schematic band diagrams for the (a) rutile, (b) M1, and (c) M2 forms of $VO_2$ according to the Goodenough picture [18,21].

One important notice has been given on the fact that the splitting of the $d_{//}$ band experimentally observed is too large to explain only by the lattice distortion [111]. Then, it is necessary to involve the effect of electron correlations [24]: the splitting of ~2.5 eV between the bonding and antibonding $d_{//}$ bands may be a sum of two energies; $D$ from dimerization and $U$ from electron correlations. Electron correlations in the rutile phase have been evidenced by the observation of enhanced Pauli paramagnetic spin susceptibility [46,112]. However, they are smaller in rutile

VO$_2$ than in V$_2$O$_3$ [3] and are eventually enhanced in the M1 phase because of two factors: screening provided by $\pi^*$ electrons has gone and the $d_{//}$ band has narrowed owing to the lack of mixing with the $\pi^*$ states [24,109]. Note that this enhancement in electron correlations is definitely triggered by the structural deformation: the additional splitting of the $d_{//}$ band by $U$ is just the consequence of the upshift of the $\pi^*$ band associated with the antiferroelectric shift of V atoms. Electron correlations in rutile VO$_2$ may not be strong enough to cause a Mott transition by itself.

Another issue to support the role of electron correlations is the presence of the M2 phase [7,24]. The M2 phase has been considered as a Mott–Hubbard insulator, because the uniform, zigzag V$^{4+}$ chain of the M2 phase (Fig. 16) behaves magnetically as a spin-1/2 Heisenberg antiferromagnetic chain. In addition, the M1 phase is also assumed be a Mott–Hubbard insulator, because it is supposed to be a superposition of two M2-type lattice distortions [7]. The electronic structure of the M2 phase can be approximately described as depicted in Fig. 18(c), following the Goodenough picture: the $d_{//}$ band of chain A with dimerization has a splitting of $D' + U'$ and that of chain B without dimerization has a smaller splitting of $U'$. Thus, to make the M2 phase insulating, electron correlations in chain B are indispensable. Nevertheless, this does not mean that electron correlations drive the transition, because this additional splitting by $U'$ in chain B must have shown up by the upshift of the $\pi^*$ band above $E_F$ as a result of the structural transition between the R and M2 phases, as in the R–M1 transition. By the way, if electron correlations give the driving force, why $T_{MI}$ is much higher in NbO$_2$ with less electron correlations than in VO$_2$?

Next, how about the Peierls instability? There is no apparent one-dimensional feature in the band structure and no associated nesting in rutile VO$_2$ [22]. Moreover, the resistivity along the $c$ axis is even larger than that perpendicular to the $c$ axis [113]. It has been claimed, however, that a Peierls instability is hidden [23] or that orbital switching makes the system more one-dimensional and thereby susceptible to a Peierls-like transition near a Mott insulating regime in the so-called Mott–Peierls scenario [27]. It is important to point out here that in a transition driven by any FS instability, such as a charge-density wave transition, an accompanying lattice deformation is always small: it is in fact negligibly small in the charge-density wave transition of a quasi-one-dimensional compound NbSe$_3$ [114]. This is because an energy gain associated with an FS instability is small, as only a small number of electrons near $E_F$ can contribute. In sharp contrast, the lattice deformations in VO$_2$ and NbO$_2$ are huge, suggesting that not only electrons near $E_F$ but also dominant bonding electrons far below $E_F$ are responsible for the stabilizations. Hence, it is not realistic to assume a Peierls mechanism. Wentzcovitch and coworkers show in their band structure calculations that the M1 structure is reproduced by the local-density approximation, though the band gap is underestimated as usually the case for such a calculation [6]. Eyert also shows this for the M2 phase [23]. Then, they claim that the dimerization is connected to the band picture rather than electron correlations. Nevertheless, the fact that the M1 and M2 structures are reproduced properly by band-structure calculations does not mean that the origin of the MIT comes from an FS instability.

In contrast, there are many data showing the importance of the structural distortion in the MITs of VO$_2$. Here I just refer recent state-of-the-art experiments to show the primary role of structural components. Cavalleri and coworkers carried out ultrafast spectroscopy experiments using femtosecond x-ray and visible light pulses and found evidence for a structurally-driven MIT in VO$_2$ [9,115]: the MIT is delayed with respect to hole injection by photo-doping, exhibiting a bottleneck time scale associated with the phonon connecting the two structural phases. The importance of the structural transitions has also been recognized by ultrafast electron diffraction experiments [116].

*2.5.2. Summary of implications in the present study*

Let me summarize the major observations given in the present study on the MIT of VO$_2$:

1. The MIT is robust up to 20% Ti$^{4+}$ substitution for V$^{4+}$ (hole doping), and orphan spins from Ti$^{4+}$–V$^{4+}$ dimers appear in the insulating phase.
2. It occurs even in thin lamellas of less than 20 Å thick in spinodally-decomposed samples.
3. There is an inherent structural instability in the rutile structure leading to the formation of metal–metal bondings, and both VO$_2$ and NbO$_2$ in the rutile form exist on the verge.
4. Stabilization of distorted structures by forming metal–metal bondings is a ubiquitous phenomenon for early transition metal compounds comprising strands of edge- or face-sharing octahedra.

The robustness of the MIT in observations 1 and 2 may be difficult to understand in terms of either electron correlations or FS instabilities. One would expect a drastic suppression of the MIT by changing band filling in such a wide range and by disorder caused by the dilution, if the transition was governed by them. The fact that the MIT is not influenced by shortening or cutting chains into short pieces (Fig. 6) implies that one-dimensionality is not crucial and also that the MIT essentially takes on a local character. Related to this, it is meaningful to refer that an MIT was observed even in three-atom-thick VO$_2$ layers in a TiO$_2$/VO$_2$ superlattice film [117].

Note that the MITs of Ti$_4$O$_7$ and V$_4$O$_7$, which occur in four-octahedra thick rutile slabs, have a great similarity with that of VO$_2$; it would be unrealistic to assume different mechanisms for these MITs. Since the electron fillings are quite different, $d^1$, $d^{0.5}$ and $d^{1.5}$ for VO$_2$, Ti$_4$O$_7$ and V$_4$O$_7$, respectively, the common mechanism would be independent of the details of electronic structures.

Observations 3 and 4 illustrate the important role of the structural instability toward a metal–metal bonding, which is common to VO$_2$ and NbO$_2$, and may be also to the Magnéli phases with MITs. I will discuss this in more detail in the following paragraphs.

*2.5.3. Structural instability driving the MITs*

Here, I would like to emphasize a more vital role of structural transitions than expected for the Peierls transition. As pointed out by the previous and present crystal chemistry considerations on the rutile family of compounds, both $VO_2$ and $NbO_2$ are located on the border between the rutile and distorted structures [67,81]. The two compounds in the rutile form already exhibit certain deviations from the structural trends for the rutile compounds. Therefore, in these compounds, even a small perturbation can trigger a structural transition between the two polymorphs. One possible trigger for the transitions is found in the temperature dependence of the $c/a$ ratio shown in Fig. 19. As $c$ decreases more quickly than $a$ upon cooling in the rutile structure as observed in $TiO_2$ [92], the $c/a$ ratio becomes small at low temperatures in $VO_2$ [74] and $NbO_2$ [29]. Then, the distorted structures are going to be selected at low temperatures. The reduction of $c/a$ upon cooling means a tendency toward a breaking of the covalent O–O bonding which increases $d(O-O)_1$ and opens a channel to allow a direct metal–metal bonding. Since all the $d$ electrons of $VO_2$ and $NbO_2$ can be accommodated in the low-lying $d_{//}$ band or be captured by metal–metal dimers in the distorted structures, the structural transitions are to be accompanied by MITs [118].

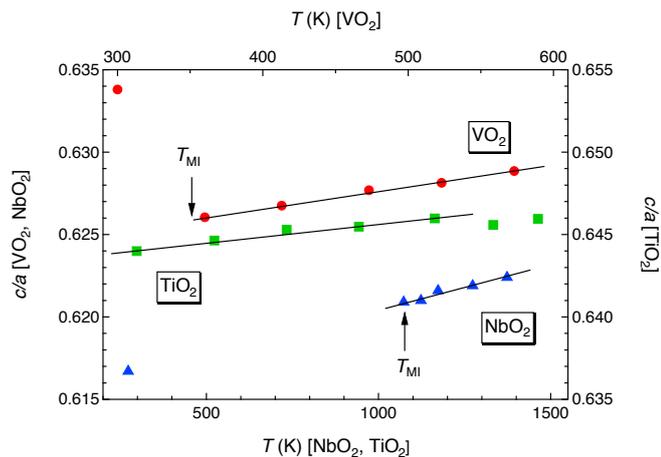

Fig. 19. Temperature dependences of $c/a$ for $VO_2$ [74], $NbO_2$ [29], and $TiO_2$ [92].

In a general sense, a structure with lower symmetry is preferred at low temperatures because of the entropy term in Gibbs free energy, so that a transition to a lower-symmetry structure can naturally occur upon cooling, if the accompanying increase in lattice energy is relatively small. On one hand, Adler and Brooks considered in a general scheme that a compound with a half-filled narrow band prefers a low-symmetry structure with a band gap of the same size as the bandwidth at low temperatures [108]. Upon heating a thermal excitation of electrons from the bonding valence band to the anti-bonding conduction band occurs and tends to stabilize a high-symmetry phase so as to reduce the gap, because the excited electrons do not contribute to the chemical bonding and thus reduce the lattice distortion [119]. This results in a first-order phase transition in the case of a narrow band, while, in the case of a wide band, a gap can be generated by antiferromagnetic order without a lattice distortion, which results in a second-order transition [108,120]. The former case may be close to the situations of $VO_2$ and $NbO_2$. However, their MITs are more strenuous than expected in such a general scheme, and a specific cause must be considered for them.

The most important question is what is the origin of the structural instability causing the structural transitions and accompanying MITs in $VO_2$ and $NbO_2$. I think that it is related to the formation of a metal–metal bonding: the distorted structures are stabilized when an energy gain by generating a metal–metal bonding is more than make up for a loss in the lattice energy. Let's go back to the $a$–$c$ diagram of Fig. 9, which is also presented in a different manner as a function of $c/a$ in Fig. 20, emphasizing the difference in the electronic property of compounds. The presence of the two $d^1$ compounds at the borderline of $c/a = 0.625$ is not just a coincidence. One expects a larger energy gain by forming metal–metal bonding for $d^2$ and $d^3$ electron systems particularly in the case of extended $4d$ or $5d$ orbitals, because the bonding states of the $t_{2g}$ orbitals in a dimer can be occupied by three electrons from each atom at maximum in the simple dimer model. In fact, the bond order estimated from the metal–metal distance is 1 for $VO_2$ and $NbO_2$, while 2 for $MoO_2$, $TcO_2$, $WO_2$ and α-$ReO_2$ [67,99]. This simple expectation has been rationalized by elaborate electronic structure calculations: Burdett has compared the rutile and $MoO_2$-type structures as a function of the $d$-electron number and found that the $MoO_2$-type structure is more stable for $d^2$ and $d^3$ than $d^1$ and less stable for $d^4$ and $d^5$ as additional electrons occupy the antibonding states [35]. Therefore, $MoO_2$, $TcO_2$, $WO_2$ and α-$ReO_2$ are located in the left of the borderline with small $c/a$ values in Fig. 20, while $RuO_2$, $RhO_2$, $OsO_2$ and $IrO_2$ appear far to the right with large $c/a$ values. The two kinds of structure are eventually in competition with each other only at the $d^1$ compounds of $VO_2$ and $NbO_2$ near the border, so that even a small perturbation can trigger the structural transitions between the two structures. $TaO_2$ with the $5d^1$ configuration is another candidate for an MIT in the Goodenough model, but seems to stay in metal, as the structural instability is weak judging from its $c/a$ value considerably larger than 0.625. This is possibly because the bonding $d_{//}$ band hybridizes with the π* band owing to the extended nature of the $5d$ orbitals, so that the energy gain by the dimerization may not be large enough for $d^1$ compared with $d^2$ or $d^3$. Note that $MoO_2$, $TcO_2$, $WO_2$ and α-$ReO_2$ in the $MoO_2$-type structure show metallic conductivity because of the partially filled π* band: the dimerization is mainly associated with the splitting of the $d_{//}$ band, and the structural boundary is nothing to do with the metal–insulator boundary, as shown in Fig. 20.

It would be interesting to test what would happen if one could control the $c/a$ ratio on some compounds. Applying pressure may change the cell volume but not the $c/a$ ratio so much; a range to be experimentally accessible may be small. Alternatively, band structure calculations would be helpful to test the idea. For example, as a function of $c/a$, $TaO_2$ may take a distorted structure at $c/a < 0.625$, which should be accompanied by an MIT, while $MoO_2$ might prefer an undistorted structure at large $c/a$ with keeping its metallic nature.

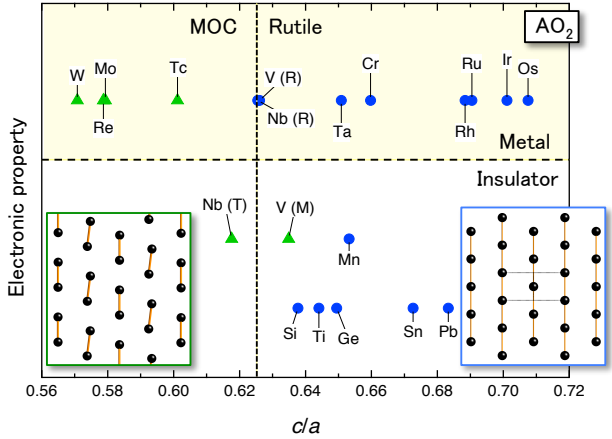

Fig. 20. Rutile-type metal dioxides lined up as a function of $c/a$ and in terms of electronic property. Located in the right and left of the vertical broken line at $c/a = 0.625$ are compounds with the rutile and distorted structures comprising dimer molecules [molecular orbital crystals (MOCs)], respectively. The top and bottom halves include metallic and insulating compounds, respectively.

The present structural instability should not be related to a Peierls or an FS instability, because the dimer structures are more stable at the metallic compounds with the $d^2$ and $d^3$ electron configurations. If it coupled with any electronic instability between metallic and insulating phases, the dimer structures would occur only near the boundary between them or only in the insulating phases. I mean that an electronic stabilization by folding a band and opening a gap at $E_F$ is not a driving force towards the dimer structures. Moreover, it is unlikely that electron correlations that are basically repulsive and weak for $4d$ and $5d$ elements can stabilize the dimer structures. The fact that $T_{MI}$ is much higher in $NbO_2$ than in $VO_2$ comes from stronger pairing for the Nb–Nb bond than the V–V bond; the $4d$ orbitals of Nb are more expanded than the $3d$ orbitals of V to form a stronger metal–metal bond.

One may claim that electron–phonon interactions are crucial for the structural instability. Related to this, Chakraverty considered a bipolaron [38,42]. However, the terminology of bipolaron implicitly assumes itinerant electrons that interact strongly with lattices and are dressed up with phonons near the metal–insulator boundary, which is obviously not the cases for $MoO_2$ and other metallic compounds. The fact that the dimerization is even stronger in these metallic compounds than in insulating $VO_2$ rules out a dominant contribution of electron–phonon interactions on the mechanism of the pairing, because screening effects by conduction electrons should weaken electron–phonon interactions and thus tend to remove distortions in metallic compounds.

On one hand, our observations reveal that the MIT of $VO_2$ is robust against the Ti substitution with keeping the dimer structure when the V chains are cut into short pieces including metal pairs giving orphan spins. In addition, an MIT is observed at ~300 K even in three-atom-thick $VO_2$ layers in a $TiO_2$/$VO_2$ superlattice film [117]. These evidence a strong local character of the stabilization of the dimer structure, implying that a classical molecular orbital picture is more adequate than a modern extended electron picture to describe the MIT. This is the reason why the simple molecular orbital picture by Goodenough works so well to explain many experimental results on $VO_2$. The driving force to the dimer phases, insulating phases in the cases of $VO_2$ and $NbO_2$ and metallic phases for $MoO_2$ and others, must be the generation of local homopolar bonds [14,15].

Plausibly, the resulting states containing dimer molecules must be robust against atomic substitutions, but the associated energy gain may be reduced by dilution. The fact that the MIT and the structural transition of $VO_2$ disappear above ~20% substitution, almost irrespective of substitution elements, suggests that there is a critical density of V–V dimer molecules to induce a cooperative structural transition. In addition, the fact that the MIT occurs in three-atom-thick layers means that one V pair is enough along the $c$ axis for the cooperative transition, while much more pairs may be required along the basal plane; this is the reason why the MIT of $VO_2$ can occur in very thin films and thus is useful in technological applications. Neither the presence of "chains" nor electronic instability in the $k$ space is the essential requirement for the transition, and critical is that more than a certain number of V atoms can find their partners to be fallen into dimer molecules so as to gain enough chemical bonding energy in total.

### 2.5.4. Substitution effects

Next, I would like to address the substitution effects on the $T_{MI}$ of $VO_2$ shown in Fig. 2. The effects may have a twofold significance: a change in electron filling and an influence on the structural stability. The former may cause a reduction of $T_{MI}$ and the latter can be complex. The decreases of $T_{MI}$ for Nb, Mo and W are natural from the viewpoint of electron filling, while the increases of $T_{MI}$ for Ti and Cr are unusual, which must be due to the introduction of another M2 phase between the R and M1 phases: the M2 phase is stabilized with the substitutions and the $T_{MI}$ at the R–M2 structure transition increases. The substitutions may prefer a uniform chain, so as to stabilize a compromise structure having both uniform and dimerized chains in the M2 phase before reaching the R phase with all uniform chains. Presumably, a cooperative dimerization becomes difficult by a disorder effect with increasing alien elements with different electron configurations. In fact, $VReO_4$ with a random distribution of V and Re atoms prefers the undistorted structure, though it is located below the borderline in Fig. 9.

The reappearance of MITs for the W substitution above 10% is mysterious [56,57]. Although Shibuya and coworkers discussed this in terms of a competition between the Mott and Peierls instabilities, it is difficult for me to believe that the two insulating states have different electronic origins. Possibly, there is another type of structural instability in the W-rich side. Note that a threefold superstructure having $V^{3+}$–$V^{3+}$ ($V^{4+}$–$V^{4+}$) pairs separated by $W^{6+}$ ($W^{4+}$) ions occurs in the trirutile structure of $V_2WO_6$ [121,122]. Thus, one would expect a structural transition from a uniform V–V–W chain to a dimerized (V–V)–W chain in $V_2WO_6$, which must be accompanied by an MIT. The V substitution for W in $V_2WO_6$ may suppress the transitions, as in the W substitution for V in VO$_2$, because extra V atoms at the W site tend to disturb a coherent V–V pairing in such a chain as W–(V–V)–V–(V–V)–W. Thus, I believe that this mystery on the W substitution is also understood within the structural instability scenario of another kind towards dimer phases.

*2.6. Molecular orbital crystals*

*2.6.1. Dimer crystals*

I would like to discuss the distorted structures of the rutile-related oxides in more detail together with those of related halides such as MoBr$_3$ and NbCl$_4$. The common origin of the structural distortions from the corresponding actual or virtual regular structures is obviously the dimerization of metal ions by forming a strong metal–metal bonding. In the case of halides, it seems that there are no phase transitions to corresponding regular structures with evenly-spaced metal atoms at high temperatures; MoBr$_3$ and RuBr$_3$ show structural transitions at 466 and 377 K, respectively, to a high-symmetry form having a random distribution of two types of dimerized chains with different phases along the chain axis [103]. Here, other types of structures having already segmented pairs of polyhedron, such as those of V$_2$O$_3$, La$_4$Re$_6$O$_{19}$ [99], or La$_3$Ru$_3$O$_{11}$ [123], are not considered.

In an actual crystal, a competition between the electronic energy gain and the loss in lattice energy, both caused by dimerization, is critical for choosing regular or distorted structures. The electronic energy gain by forming a metal–metal bonding should depend on the number of $d$ electrons in the $t_{2g}$ orbitals. When a pairing occurs in an edge-sharing octahedron chain, as schematically depicted in Fig. 21, three kinds of molecular orbitals (MOs) are generated in a metal pair: the largest stabilization is attained in the $\sigma$ MO from the $d_{xy}$ orbitals, and from the $d_{yz}/d_{zx}$ orbitals weakly bonding $\pi$ and almost nonbonding $\delta$ MOs are formed [124]. Thus, $d^1$ and $d^2$ systems like NbCl$_4$ and WCl$_4$ naturally favor distorted structures, while the choice becomes delicate for systems with more electrons. In fact, a regular structure is attained for OsCl$_4$ with a $d^4$ configuration [105]. In addition, note that the total electronic energy must be determined by a balance between the removal of chemical bonds already present in a regular structure and the emergence of new metal–metal bonds in a distorted structure. For example, in rutile VO$_2$, there is a metallic bond from $d_{xy}$ orbitals via oxygen $2p$ orbitals along the chain, while it is replaced by a set of direct metal–metal bonds on every second V–V pairs in its distorted structure; the latter gains more bonding energy than the former so that the structural transition takes place.

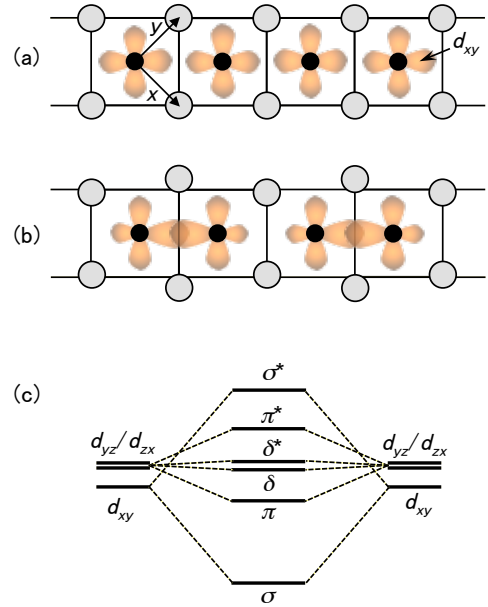

Fig. 21. Schematic drawings of (a) a regular chain made of edge-sharing octahedra, (b) a distorted chain with cation pairs, and (c) the energy levels of molecular orbitals for a cation pair. The black and grey circles in (a) and (b) represent transition metal cations and anions, respectively. Bridging anions between paired cations shift outward to allow a direct overlap between the $t_{2g}$ orbitals. The clovers represent $d_{xy}$ orbitals which lie in the plane and heavily overlap with each other in a metal pair, giving the $\sigma$-type bonding molecular orbital in (c). The other $d_{yz}$ and $d_{zx}$ orbitals give weakly bonding $\pi$ and almost nonbonding $\delta$ molecular orbitals. A small splitting between the $d_{xy}$ and $d_{yz}/d_{zx}$ levels has been assumed, as the AX$_6$ octahedron is compressed perpendicular to the paper in many compounds.

On the other hand, the lattice energy has two aspects: one is the electrostatic energy, which favors a regular octahedron with a cation at the center of gravity of surrounding anions, and the other is the space filling. In general, the crystal structures of transition metal compounds are considered to be built up by putting small cations into voids among close-packed large anions. Then, the packings of anions are slightly relaxed in various ways so as to adjust the size of voids to fit the cations and also to stabilize directional $d$ orbitals in an anion polyhedron. Therefore, it is always important to find an enough space for large anions to move as a result of distortions. In the present dimerization, it must be critical to find reasonable ways to accommodate the large lateral movements of bridging anions between coupled cations such as schematically depicted in Fig. 21.

Since the cooperative dimerization is quite ubiquitous among compounds having chain-like units made of face- or edge-sharing AX$_6$ octahedra, the associated lattice energy loss is potentially small compared with an energy gain by forming metal–metal bonds in these structures. In the case of halides, the strands of octahedra are isolated from each other, as in MoBr$_3$ and NbCl$_4$ in Fig. 17. Thus, the lattice deformations, particularly the large lateral movements of bridging anions, can be easily absorbed by minimal interference with neighboring strands. This may not be easy in structures that lack such a one-dimensional character; more complex ways to avoid an increase in the total lattice energy in terms of space filling would be required. In the case of the rutile structure, strands of octahedra are not isolated but interconnected to each other by common vertices. Nevertheless, thanks to the flexibility of the rutile structure, the large lateral movements of bridging O ions are well accommodated in smart ways as described for the M1 and M2 structures in Figs. 15 and 16, respectively. Therefore, the frequent appearance of dimerization in metal chains is just a consequence of advantage to align dimers one-dimensionally with a small increase in lattice energy in actual three-dimensional structures, which is absolutely not related to the Peierls instability.

I would like to call this class of compounds with all or a part of transition metal ions occurring in pairs at much reduced bond lengths by forming direct metal–metal bonds "dimer crystals". As discussed in the preceding paragraph, the driving force to dimer crystals is nothing to do with FS instabilities; it is only required to have enough electrons in the bonding orbitals irrespective of the density of conduction electrons. Thus, the electronic properties of dimer crystals can be insulating as in $VO_2$, $NbO_2$, the Magnéli phases, and the halides mentioned above, or metallic as in $MoO_2$ and others, depending on the electron fillings and the details of band structures.

It might be possible to call some of them "bipolaron crystals", following the idea by Chakraverty for $VO_2$ and $Ti_4O_7$ [38,42]. However, as mentioned also in the preceding paragraph, the terminology of bipolaron implicitly assumes itinerant electrons that interact strongly with lattice near a metal–insulator boundary. Although it is not impossible to consider that these compounds, only insulating ones, not metallic ones, lie in the limit of strong electron–phonon interactions, I would prefer a more realistic picture based on the dimer crystal with greater emphasis on chemical bondings than electron–phonon interactions.

Alternatively, one may prefer a terminology of valence-bond crystal (VBC), which may emphasize an order–disorder character of valence bonds. However, I would like to employ the concept of the dimer crystal for these compounds, because the key issue is the generation of new chemical bonds in the dimer crystal that have been missing in the corresponding regular structure; the enthalpy term in free energy is dominant rather than the entropy term.

The MITs of $VO_2$, $NbO_2$, and the Magnéli phases are transitions to insulating dimer crystals, neither to Mott–Hubbard nor Peierls insulators. It is again emphasized here that the key feature of the dimer crystal is an assembly of local dimer molecules, so that it is robust against dilution and can occur even in atomically thin layers.

### 2.6.2. Molecular orbital crystals

Expanding this idea of the dimer crystal one step further, one conceives of a terminology of "molecular orbital crystal (MOC)" that assumes an array of "molecules" made of transition metal atoms, a dimer or larger ones, embedded in a crystal. Each molecule is locally stabilized by a metal–metal bonding or a direct overlap among $d$ orbitals with its bonding MOs occupied by enough pairs of electrons. Note that the molecule is generated by a relatively weak chemical bond compared with much stronger ones that build a rigid framework of the crystal, just as in organic molecules having $\pi$ and $\sigma$ bonds. Candidates for the MOC are listed in Table 5, including the rutile-related compounds mentioned thus far.

Typical examples of the MOC are found in complex molybdenum oxides. Trinuclear $Mo_3$ clusters were first discovered in $Zn_2Mo_3O_8$ ($Mo^{4+}$; $4d^2$) [125], and are now known in a wide range of compounds [126-128]. As depicted in Fig. 22(a), $Zn_2Mo_3O_8$ possesses a two-dimensional network of Mo atoms placed in oxygen octahedra connected by their edges, which consists of two kinds of Mo triangles, one with a short Mo–Mo distance of 2.524 Å and the other with a long distance of 3.235 Å (25% difference) [126,127]. Thus, the Mo network is considered as a "breathing" kagome lattice. It is clear that this deformation of the kagome lattice is due to the formation of Mo timers having direct metal–metal bonds [128]. $LiZn_2Mo_3O_8$ ($Mo^{3.67+}$; $4d^{2.33}$) comprises similar Mo trimers with a ~23% reduced bond length [127,129].

Another examples of the MOC are found in the family of hollandite compounds. $K_2Mo_8O_{16}$ ($Mo^{3.75+}$; $4d^{2.25}$) crystallizes in the hollandite structure consisting of double strands of edge-sharing $MoO_6$ octahedra interconnected via corner sharing [Fig. 22(b)] [130]. In the double strands, planer tetranuclear clusters are formed, which are stabilized by Mo–Mo direct bondings with 9 electrons in the cluster bonding orbitals. The difference in the Mo–Mo bond length reaches about 20%, as large as in $MoO_2$, which makes it easy and meaningful to identify the "molecules" in the lattice. The electronic state of $K_2Mo_8O_{16}$ has been approximated as a solid of "superatoms" of the Mo tetramers, and thus is considered as an MOC in analogy of molecular crystals made of organic molecules [131]. Similar Mo tetramers are observed in a related compound $Rb_{1.5}Mo_8O_{16}$, which shows a semiconductor-like resistivity governed by cluster-to-cluster electron hopping at high temperatures and becomes more insulating below 208 K [132].

There are not a few transition metal compounds other than the molybdenum oxides to be referred in this context, most of which exhibit phase transitions accompanied by MITs or insulator–insulator transitions (IITs) to lower-symmetry structures upon cooling. For example, an IIT is observed at 540 K in $Li_2RuO_3$ ($Ru^{4+}$; $4d^4$), where the high-temperature structure with a slightly-distorted honeycomb lattice made of $Ru^{4+}$ ions is heavily deformed so as to generate Ru–Ru dimers with ~16% reduced bond lengths at low temperatures [Fig. 22(c)] [124]. It has been pointed out that this transition is driven by the formation of molecular orbitals in the dimers [133]. The peculiar ordering pattern of dimers in the honeycomb net seems to be explained by taking into account a magnetoelastic coupling [134].

Many vanadium and titanium oxides are known to exhibit MITs or IITs accompanied by the formation of clusters of transition metals. V trimers are found in two triangular-lattice compounds with $V^{3+}$ ($3d^2$); $LiVO_2$ below 490 K [Fig. 22(d)] [135] and $LiVS_2$ below 300 K [136]. A similar V trimerization is observed in the complex three-dimensional lattice of $BaV_{10}O_{15}$ below 130 K [137,138]. In the hollandite vanadate $K_2V_8O_{16}$ ($V^{3.75+}$; $3d^{1.25}$), three $V^{4+}$– $V^{4+}$ dimers and one $V^{3+}$–$V^{3+}$ dimer seem to appear below an MIT at 170 K [139]. The two pyroxene-type $Ti^{3+}$ ($3d^1$) compounds, $NaTiSi_2O_6$ and $LiTiSi_2O_6$, can be another interesting examples for the MOC, which crystallize in a structure having chains made of skew edge-sharing $TiO_6$ octahedra [140]. They show IITs at 210 and 230 K, respectively, from quasi-one-dimensional antiferromagnets to dimerized, spin-singlet magnets upon cooling; a 7% bond alternation is observed at 10 K in $NaTiSi_2O_6$ [141]. The origin has been ascribed not to a spin–Peierls transition but a sort of orbital ordering [140,142,143]. Provided that a gain in the chemical bonding energy by forming the V or Ti clusters is dominant, the low-temperature phases of these vanadates and titanates could be classified into MOCs.

There are more related compounds in the family of spinel compounds that show similar transitions. For example, dimerizations and associated MITs or IITs are observed at 260 K in $MgTi_2O_4$ ($Ti^{3+}$; $3d^1$) [144] and at 170 K in $LiRh_2O_4$ ($Rh^{3.5+}$; $4d^{5.5}$) [145]. Moreover, the low-temperature phases of $AlV_2O_4$ ($V^{2.5+}$; $3d^{2.5}$) below 700 K and $CuIr_2S_4$ ($Ir^{3.5+}$; $5d^{5.5}$) below 230 K comprise heptamers [146] and octamers (or four-dimer complexes) [147], respectively, in the three-dimensional pyrochlore lattices of transition metal atoms. On the other hand, a pressure-induced MIT has been observed in the heavy fermion compound $LiV_2O_4$ ($V^{3.5+}$; $3d^{1.5}$) above 9 GPa [148], where the formation of vanadium molecules such as found in $AlV_2O_4$ is suggested [149], though the crystal structure has not been determined.

I am not so sure whether these spinel compounds are to be classified into MOCs or not. One reason is that the resulting bond alternations, which are not always available owing to difficulty in structural analyses, may be relatively small in some compounds, compared with those of $VO_2$ and other examples mentioned above, which is likely due to the rigid three-dimensional frameworks. If an order–disorder nature is more important for these transitions, orbital order and resulting Peierls instability could be one explanation, as pointed out by Khomskii and Mizokawa for $MgTi_2O_4$ and $CuIr_2S_4$ [150]. Then, the charge/orbital ordering transition might be an appropriate terminology. Nevertheless, I think that the molecular orbital crystallization plays a crucial role in these spinel compounds too, because there must be a considerably large energy gain by generating molecular orbitals from the partially-filled $t_{2g}$ orbitals. Magnetite $Fe_3O_4$ having the spinel structure can be another good example for the MOC, as will be addressed in the next section.

Concerning the relation between the orbital order and the MOC, it is meaningful to focus on the successive transitions found by Okamoto and coworkers in $LiRh_2O_4$ in more detail [145]. The compound first exhibits an orbital ordering transition or a band Jahn–Teller transition at 230 K upon cooling by selecting the $d_{xy}$ orbital out of the triply degenerate $t_{2g}$ orbitals, so that the structure changes from cubic to tetragonal there. Then, a second transition sets in at 170 K, where both an MIT and a dimerization into a nonmagnetic state take place simultaneously with all the itinerant $d_{xy}$ electrons trapped into direct Rh–Rh bonds. This second transition looks quite similar to the MIT of $VO_2$, as the partially-filled band in the chain of $d_{xy}$ orbitals (the $d_{//}$ band for $VO_2$) is transformed into the row of metal dimers in the edge-sharing octahedron chain embedded in the three-dimensional spinel structure. Thus, it is definitely classified into a molecular orbital crystallization. This illustrates that an orbital ordering transition (or a Jahn–Teller transition) and an MOC transition can occur separately with different driving forces, that is, the entropy associated with orbital degeneracy (or a Jahn–Teller energy) and the enthalpy associated with chemical bonding. The reason why a single transition is observed in every other spinel compound may be that the two types of transitions happen to take place simultaneously.

As for the magnitude of lattice deformations, I have assumed a "large" bond alternation as an important condition for the MOC. However, it should not be a primary condition, because the magnitude must depend on the surrounding structures and is determined by a balance between the electronic and structural energies. Mazin and coworkers called $Na_2IrO_3$ ($Ir^{4+}$; $5d^5$) a molecular orbital crystal even in the absence of a lattice distortion, whose electronic state may be dominated by the formation of quasimolecular orbitals in the hexagon of Ir atoms [151]. The presence of molecules made of transition metals via direct metal–metal bonding is the minimum requirement for the MOC.

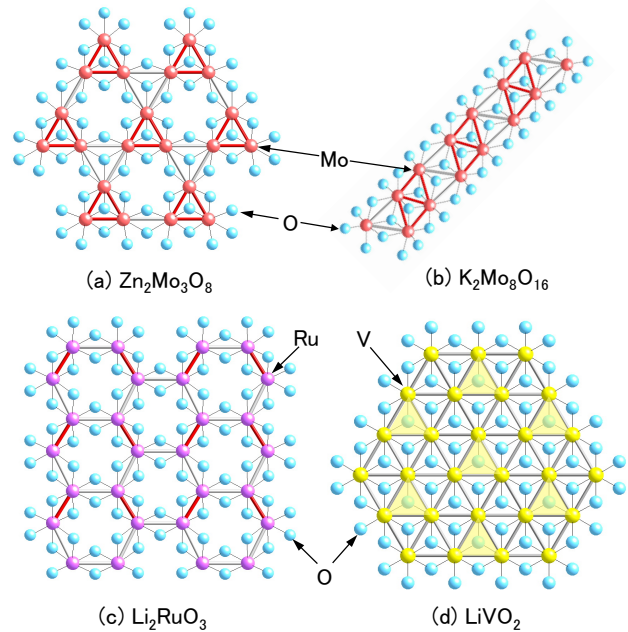

Fig. 22. Possible candidates for the molecular orbital crystal (MOC). (a) Mo trimers occur in the kagome-like network of $Zn_2Mo_3O_8$ [125]; the bond alternation reaches 25%. (b) Mo tetramers with ~16% reduced bond lengths are embedded in the double strands of edge-sharing octahedra of $K_2Mo_8O_{16}$ [130]. (c) Ru dimers with ~20% reduced bond lengths are generated below 540 K in the honeycomb network of Ru ions of $Li_2RuO_3$ [124]. (d) V trimers are present below 490 K in the triangular lattice of $LiVO_2$ [135].

### 2.6.3. Molecular orbital crystal or orbital-molecule crystal

Here I would like to address the nature of chemical bonds in the MOCs. Generally speaking, the chemical bond in a homonuclear diatomic molecule changes its character from the Heitler–London (HL) type in the $U = \infty$ limit to the molecular orbital (MO) type in the $U = 0$ limit. In the former, each of the two electrons sits in an atomic orbital of either atom, and an exchange interaction between them stabilizes the molecule in a spin-singlet state. In the latter, on the other hand, the two electrons occupy a single bonding molecular orbital that extends over the two atoms, resulting in a nonmagnetic state with paired electrons. It is reasonably expected that the chemical bond in the MOC also changes its character between the two types, depending on the magnitudes of on-site Coulomb interactions and the overlap of $d$ orbitals. For the rutile compounds, the bonding character must vary gradually from the HL type for $VO_2$ to the MO type toward $WO_2$ with decreasing $U$ and increasing the orbital overlap, as schematically shown in Fig. 23.

Very recently, Senn, Paul Attfield and coworkers have carefully studied the Verwey transition of magnetite $Fe_3O_4$, which has been another controversial issue left for over 70 years in condensed matter science [152,153]. They are successful in determining the complex superstructure appearing at low temperatures below $T_{MI} = 120$ K and find a "trimeron" as a basic building block, which is a linear, three-Fe-atom unit carrying an effective spin-1/2. Moreover, they propose that the trimeron is regarded as a kind of "orbital molecules" that are locally-coupled orbital states on two or more metal ions within an orbital-ordered solid [152-154]. In this context, the low-temperature phase of $Fe_3O_4$ is considered to be an "orbital-molecule crystal". It is striking that the two MITs of $VO_2$ and $Fe_3O_4$ are induced by generating similar "orbital molecules" with direct metal–metal interactions (strong and weak, respectively) in the edge-sharing octahedron networks.

Figure 23 sums up all the compounds mentioned above in a map in terms of $U/t$, where $U$ and $t$ refer to the magnitudes of electron correlations and electron transfers, respectively, and the number of atoms in a molecule embedded in a crystal. The $U/t$ values used for the map are just approximate, and the number of atoms in a molecule varies as 2, 3 and more to the right. The question is which is the better terminology for these compounds, the orbital-molecule crystal or the molecular orbital crystal. Apparently, the $3d$ compounds lie to the side of the orbital-molecule crystal containing HL-type clusters, as suggested for $Fe_3O_4$, while the $4d$ and $5d$ compounds lie to the side of the MOC containing MO-type clusters. Note, however, that actual compounds should always exist between these two extreme pictures; even a hydrogen molecule takes both HL and MO characters. For the rutile compounds, the MOC picture is probably more appropriate, because the $T_{MI}$ is higher in $NbO_2$ with less electron correlations than in $VO_2$, and also because the distorted structures become more stable toward the $5d$ metal dioxides. For the other compounds, I would also like to use the MOC as a more general concept than the orbital-molecule crystal, because the chemistry and physics they show are essentially the same as in the rutile compounds. Although there may be non-negligible contributions to the chemical bonding from electron correlations particularly in the compounds at the upper side of Fig. 23, I think that a common principle, that is the MOC, should be applied for understanding the similar properties and phenomena observed in all these transition metal compounds.

In the family of spinel oxides, there are many compounds that are apparently not MOCs. For example, $ZnV_2O_4$ ($V^{3+}$; $3d^2$) undergoes an orbital ordering transition from a cubic to a tetragonal structure at 50 K, which results in two sets of orthogonal spin chains, and then an antiferromagnetic order below 40 K [155,156]: no dimers or local clusters are formed at the lowest temperature. This is in sharp contrast to $LiRh_2O_4$, in which a similar orbital-ordering transition is followed by the formation of a dimer crystal [145]. Probably, in this typical Mott–Hubbard insulator, electron correlations are too strong to allow a metal–metal bonding with a reduced interatomic distance. It is also the case for $ZnCr_2O_4$ ($Cr^{3+}$; $3d^3$), in which structural and antiferromagnetic transitions occur simultaneously at 12.5 K [157]. In addition, $V_2O_3$ gives another example that the MOC does not occur in the presence of strong electron correlations; electron correlations are apparently weaker in rutile $VO_2$ than in $V_2O_3$ [3]. Hence, the MOC is realized under moderate or weak electron correlations, which is another reason that I prefer the MOC than the orbital-molecule crystal to specify all the compounds shown in Fig. 23.

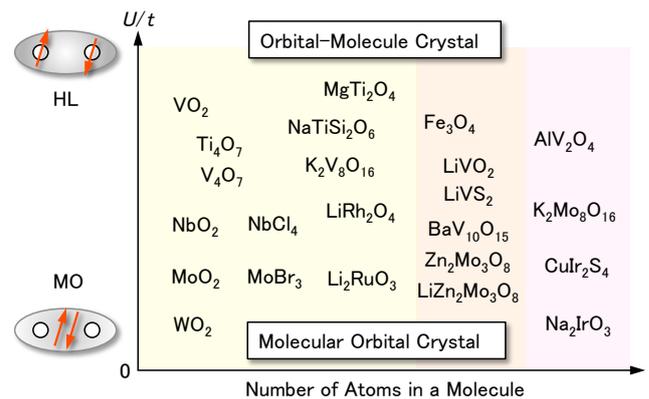

Fig. 23. Map of transition metal compounds that can be classified into MOCs in terms of $U/t$, where $U$ and $t$ refer to the magnitudes of electron correlations and electron transfers, respectively, and the number of atoms in a molecule. The $U/t$ values used are just approximate, and the number of atoms in a molecule varies as 2, 3 and more to the right. The chemical bond between atoms in a molecule changes its character from the Heitler–London (HL) type at $U/t = \infty$ to the molecular orbital (MO) type at $U/t = 0$, as schematically depicted to the left of the map. All the actual compounds in their distorted forms exist between the orbital-molecule crystal in the former limit and the molecular orbital crystal in the latter limit. All the compounds exhibit MITs or IITs except for $MoO_2$ and $WO_2$, which remain metals down to low temperatures, and $NbCl_4$, $MoBr_3$, $Zn_2Mo_3O_8$ and $LiZn_2Mo_3O_8$, which are insulators.

*2.6.4. General features of the MOCs*

There are two common features among the MOCs shown in Fig. 23 and Table 5: one is the crystal structure having $AX_6$ octahedra connected by their edges (except for $MoBr_3$), and the other is the electron configuration of transition metal ions having partially-filled $t_{2g}$ shells in the octahedral crystal field. In the corresponding undistorted structures, the former results in shorter metal–metal distances than in corner-sharing frameworks such as typically found in perovskite compounds. On the other hand, the latter gives a preference to form direct metal–metal bondings by overlapping nearby $t_{2g}$ orbitals in the distorted structures. Therefore, the MOCs are always found for early transition metal compounds in the rutile, spinel and other complex structures made of edge-sharing octahedra.

How large molecule is generated should depend on the crystal structure to find a suitable deformation pattern with a small loss in lattice energy. In addition, there is a trend to be noticed in terms of electron filling in the $t_{2g}$ shell: most $d^1$ compounds favor dimers, $d^2$ compounds favor trimers, and mixed-valence compounds near $d^3$ favor larger molecules (Table 5). This may be simply related to the number of valence electrons necessary to form a molecule by a direct metal–metal bonding: one for a dimer, two for a trimer and more for larger molecules. One exception is found in Li$_2$RuO$_3$ having dimers of $d^4$ cations, which is due to the honeycomb lattice with only three neighboring atoms. Further looking exceptional are dimer crystals with mixed valency such as Ti$_4$O$_7$ ($d^{0.5}$), V$_4$O$_7$ ($d^{1.5}$), the other Magnéli phases, and the spinel compounds LiRh$_2$O$_4$ and CuIr$_2$O$_4$ (both $d^{5.5}$). In these compounds, however, charge disproportionations are induced so that electrons are concentrated into one set of cations with the other sets having less electron density: only a set of cations having $d^1/d^5$ configurations eventually exhibit a dimerization. A similar situation may be realized in Fe$_3$O$_4$ [152,153]. The other mixed valence compounds with electron configurations between $d^2$ and $d^3$ tend to be stabilized by sharing the $d$ electrons in molecular orbitals extending over a trimer or larger ones, as typically found in the hollandite compound K$_2$Mo$_8$O$_{16}$.

As for the electronic property of the MOC, most of the compounds shown in Fig. 23 and Table 5 exhibit MITs or IITs accompanied by magnetic transitions, when the regular structures are transformed into the MOCs at low temperatures. The MIT occurs when all conduction electrons are trapped into molecular orbitals at the transition, while the IIT is achieved with energy gaps remaining all the way. Note that the IIT would be difficult to understand if one sticks to any ideas based on FS instability or electron correlations. Even a metal–metal transition may occur when conduction electrons survive through the transition, though such an example has not been known to my knowledge; MoO$_2$ could undergo such a transition, if the dimers were broken at an elevated temperature. Without a specific transition, the MOC remains a paramagnetic metal as in MoO$_2$, WO$_2$, and α-ReO$_3$ or a diamagnetic insulator as in NbCl$_4$ and MoBr$_3$.

Since the MITs of VO$_2$ and others are presumably transitions from Bloch-wave metals to insulating MOCs with the local character, the insulating states may possess electronic properties that are distinguished from those for conventional semiconductors: local excitations in the real space must prevail against band excitations in the $k$ space. In fact, the $^{51}$V NMR study on VO$_2$ by Takanashi and coworkers show that the NMR relaxation rate of the insulating phase cannot be explained by a band model and that the hopping motion of self-trapped carriers is responsible for the relaxation process [158]. The semiconductor-like resistivity of Rb$_{1.5}$Mo$_8$O$_{16}$ has also been suggested to be governed by cluster-to-cluster electron hopping [132].

The magnetic ground states of the MOCs are often nonmagnetic but not always so, as listed in Table 5. Most dimer crystals actually fall into nonmagnetic or singlet-based states, since all $d$ electrons occur in pairs. Magnetic excitations in them must take strong local characters, as typically evidenced by the appearance of orphan spins for the TiO$_2$–VO$_2$ solid solutions in Fig. 7. In contrast, a paramagnetic or an antiferromagnetic response is observed, when a part of $d$ electrons survive without pairing. For example, in V$_4$O$_7$, two sets of four-atom-long chains [Fig. 17(c)] are generated at low temperatures below $T_{MI}$ = 250 K [38,40,159]: one set of chains with V$^{4+}$ ions exhibit a dimerization to become nonmagnetic, while the other set of chains with V$^{3+}$ ions remain nearly uniform and finally undergo an antiferromagnetic long-range order below $T_{mag}$ = 40 K [39]. On the other hand, a paramagnetic state is realized when excess electrons are localized at other crystallographic sites and behave as nearly free spins with negligible couplings to the surroundings, which is actually the case for BaV$_{10}$O$_{15}$ [138] or AlV$_2$O$_4$ [134]. In addition, MoO$_2$, WO$_2$ and α-ReO$_2$ are paramagnetic metals, as a portion of $d$ electrons reside in the conduction bands [67].

When a large molecule carries odd number of electrons in its MOs, the compound is assumed to be a molecular magnet. In K$_2$Mo$_8$O$_{16}$, for example, one unpaired electron exists per Mo tetramer (9 electrons in the MOs) and behaves as a local magnetic moment. This molecular magnet remains paramagnetic down to low temperatures because of negligible interactions [120][132]. LiZn$_2$Mo$_3$O$_8$ gives an interesting system in which one electron per Mo trimer is added to the breathing kagome lattice of Zn$_2$Mo$_3$O$_8$ shown in Fig. 22(a). It seems that the compound becomes a quantum antiferromagnet with spin 1/2 per Mo trimer below 100 K and shows no magnetic order down to 50 mK, possibly because of geometrical frustration on the triangular lattice [129].

Therefore, the electronic and magnetic properties of these compounds are qualitatively interpreted in terms of the MOC: the concept of the MOC is so general that it can cover various states including metals and insulators as well as nonmagnetic, paramagnetic and antiferromagnetic states. It would be intriguing to re-examine all the accumulated experimental data on their phase transitions assuming the molecular orbital crystallization. I think that the idea of the MOC is generally important for understanding various phenomena in transition metal compounds.

## 3. Conclusions

I have reviewed the MIT of VO$_2$ through Ti substitution experiments and from a broad perspective of crystal chemistry on the rutile-related compounds. The Ti substitution experiments reveal that the MIT is robust up to 20% Ti$^{4+}$ (hole) doping and occurs even in extremely thin V-rich lamellas in spinodally decomposed $TiO_2$–$VO_2$ composites, indicating that the MIT essentially takes on a local character and suggesting that either electron correlation in the Mott–Hubbard sense or Peierls (FS) instability plays a minor role on the MIT. It is demonstrated from a crystal chemistry viewpoint that $VO_2$ and another MIT compound $NbO_2$ in the family are located just near the border between the rutile structure and the distorted structures characterized by the formation of a metal–metal bonding; the borderline having the axial ratio of $c/a$ = 0.625 separates the two structural polymorphs. The key feature of the rutile structure is the presence of a strongly covalent bond between two bridging oxide ions, which is broken to open a channel for a direct metal–metal bonding in the distorted dimer structures.

I consider that a classical molecular orbital picture is more adequate than a modern extended electron picture to grasp the true nature of the dimer phases in the rutile family. The driving force to the dimer phases must be the generation of direct metal–metal bonding. Structural transitions between the rutile and dimer phases happen to accompany MITs in $VO_2$ and $NbO_2$, as all $d^1$ electrons are accommodated in the bonding molecular orbitals of the dimers. The MITs of $VO_2$, $NbO_2$, and possibly Magnéli phases are transitions to insulating dimer crystals or molecular orbital crystals, neither to Mott–Hubbard insulators nor to Peierls insulators.

MOCs having dimer, trimer and larger molecules occur quite ubiquitously in various kinds of transition metal compounds mostly comprising edge-sharing octahedron networks and having partially-filled $t_{2g}$ shells. The molecular orbital crystallization opens one natural route to stabilization of unpaired $t_{2g}$ electrons in crystals. I believe that the concept of the MOC would provide us with a helpful basis to bring about solutions to previous controversial issues.

MITs and related phenomena in transition metal compounds have been discussed by many physicists thus far in terms of electron correlations or electronic band structures in the $k$ space. However, a chemical bonding in the real space can be more important and dominates physical properties in some compounds, where chemist intuitions are helpful.

**Acknowledgements**

I appreciate H. Hayamizu and T. Yoshida letting me use their unpublished data on the magnetic susceptibility of SSs and the resistivity of a single crystal in the $TiO_2$–$VO_2$ system, respectively. I am grateful to Y. Okamoto, J. Yamaura, A. Fujimori, R. Seshadri, and M. Subrananian for their critical reading and valuable comments on the manuscript. I also thank Y. Muraoka, Y. Ueda, H. Harima, M. Ogata, and J. Paul Attfield for fruitful discussion. Finally, let me say thank you to all readers for their patience in reading through this long story and I hope they have enjoyed it.

**References**

[1] F. J. Morin, Phys. Rev. Lett. 3 (1959) 34.
[2] N. F. Mott, Rev. Mod. Phys. 40 (1968) 677.
[3] N. F. Mott, Metal-insulator transitions, Taylor & Francis, London, New York, 1990.
[4] N. Tsuda, K. Nasu, A. Fujimori, K. Shiratori, Electronic Conduction in Oxides, Springer-Verlag, Heidelberg, 1991, 120.
[5] M. Imada, A. Fujimori, Y. Tokura, Rev. Mod. Phys. 70 (1998) 1039.
[6] R. M. Wentzcovitch, W. W. Schulz, P. B. Allen, Phys. Rev. Lett. 72 (1994) 3389.
[7] T. M. Rice, H. Launois, J. P. Pouget, Phys. Rev. Lett. 73 (1994) 3042.
[8] H. Jerominek, F. Picard, D. Vincent, Opt. Eng. 32 (1993) 2092.
[9] A. Cavalleri, C. Toth, C. W. Siders, J. A. Squier, Phys. Rev. Lett. 87 (2001) 237401.
[10] M. Gurvitch, S. Luryi, A. Polyakov, A. Shabalov, J. Appl. Phys. 106 (2009) 101504.
[11] J. Cao, E. Ertekin, V. Srinivasan, W. Fan, S. Huang, H. Zheng, J. W. L. Yim, D. R. Khanal, D. F. Ogletree, J. C. Grossman, J. Wu, Nat Nano 4 (2009) 732.
[12] M. Nakano, K. Shibuya, D. Okuyama, T. Hatano, S. Ono, M. Kawasaki, Y. Iwasa, Y. Tokura, Nature 487 (2012) 459.
[13] Z. Hiroi, H. Hayamizu, T. Yoshida, Y. Muraoka, Y. Okamoto, J. Yamaura, Y. Ueda, Chem. Mater. 25 (2013) 2202.
[14] B.-O. Marinder, A. Magnéli, Acta Chem. Scand. 11 (1957) 1635.
[15] J. B. Goodenough, Phys. Rev. 117 (1960) 1442.
[16] Y. Bando, K. Nagasawa, Y. Kato, T. Takada, Jpn. J. Appl. Phys. 8 (1969) 633.
[17] A. Magnéli, G. Andersson, Acta Chem. Scand. 9 (1955) 1378.
[18] J. B. Goodenough, J. Solid State Chem. 3 (1971) 490.
[19] M. Marezio, D. B. McWhan, J. P. Remeika, P. D. Dernier, Phys. Rev. B 5 (1972) 2541.
[20] J. B. Goodenough, Magnetism and the Chemical Bond, Interscience Publishers, Inc., New York, 1963.
[21] J. B. Goodenough, Prog. Solid State Chem. 5 (1971) 145.
[22] M. Gupta, A. J. Freeman, D. E. Ellis, Phys. Rev. B 16 (1977) 3338.
[23] V. Eyert, Ann. Phys. 11 (2002) 650.
[24] A. Zylbersztejn, N. F. Mott, Phys. Rev. B 11 (1975) 4383.
[25] D. Paquet, P. Leroux-Hugon, Phys. Rev. B 22 (1980) 5284.
[26] S. Biermann, A. Poteryaev, A. I. Lichtenstein, A. Georges, Phys. Rev. Lett. 94 (2005) 026404.
[27] M. W. Haverkort, Z. Hu, A. Tanaka, W. Reichelt, S. V. Streltsov, M. A. Korotin, V. I. Anisimov, H. H. Hsieh, H. J. Lin, C. T. Chen, D. I. Khomskii, L. H. Tjeng, Phys. Rev. Lett. 95 (2005) 196404.
[28] R. F. Janninck, D. H. Whitmore, J. Phys. Chem. Solids 27 (1966) 1183.
[29] T. Sakata, Phys. Stat. Solidi. B 20 (1967) K155
[30] K. Sakata, J. Phys. Soc. Jpn. 26 (1969) 582.
[31] V. Eyert, EPL 58 (2002) 851.
[32] G. Andersson, Acta Chem. Scand. 8 (1954) 1599.
[33] S. Andersson, B. Collen, U. Kuylenstierna, A. Magneli, Acta Chem. Scand. 11 (1957) 1641.
[34] R. F. Bartholomew, D. R. Frankl, Phys. Rev. 187 (1969) 828.
[35] J. K. Burdett, Inorg. Chem. 24 (1985) 2244.
[36] M. Marezio, D. McWhan, P. Dernier, J. Remeika, Phys. Rev. Lett. 28 (1972) 1390.
[37] M. Marezio, D. B. McWhan, P. D. Dernier, J. P. Remeika, J. Solid State Chem. 6 (1973) 213.
[38] S. Lakkis, C. Schlenker, B. K. Chakraverty, R. Buder, M. Marezio, Phys. Rev. B 14 (1976) 1429.

[39] S. Kachi, K. Kosuge, H. Okinaka, J. Solid State Chem. 6 (1973) 258.
[40] M. Marezio, D. B. McWhan, P. D. Dernier, J. P. Remeika, J. Solid State Chem. 6 (1973) 419.
[41] M. Marezio, P. D. Dernier, D. B. McWhan, S. Kachi, J. Solid State Chem. 11 (1974) 301.
[42] B. K. Chakraverty, J. Mag. Mag. Mater. 3 (1976) 176.
[43] B.-O. Marinder, A. Magnéli, Acta Chem. Scand. 12 (1958) 1345.
[44] G. Villeneuve, A. Bordet, A. Casalot, P. Hagenmuller, Mater. Res. Bull. 6 (1971) 119.
[45] J. B. Goodenough, H. Y. P. Hong, Phys. Rev. B 8 (1973) 1323.
[46] J. P. Pouget, H. Launois, T. M. Rice, P. Dernier, A. Gossard, G. Villeneuve, P. Hagenmuller, Phys. Rev. B 10 (1974) 1801.
[47] M. Drillon, G. Villeneuve, Mater. Res. Bull. 9 (1974) 1199.
[48] K. Kosuge, J. Phys. Soc. Jpn. 22 (1967) 551.
[49] W. Rüdorff, G. Walter, J. Stadler, Z. Anorg. Allg. Chem. 297 (1958) 1.
[50] T. Hörlin, T. Niklewski, M. Nygren, Acta Chem. Scand. A 30 (1976) 619.
[51] T. Hörlin, T. Niklewski, M. Nygren, J. Phys. Colloque C4 (1976) 69.
[52] K. Sakata, T. Sakata, Jpn. J. Appl. Phys. 6 (1967) 112.
[53] G. Villeneuve, A. Bordet, A. Casalot, J. P. Pouget, H. Launois, P. Lederer, J. Phys. Chem. Solids 33 (1972) 1953.
[54] K. L. Holman, T. M. McQueen, A. J. Williams, T. Klimczuk, P. W. Stephens, H. W. Zandbergen, Q. Xu, F. Ronning, R. J. Cava, Phys. Rev. B 79 (2009) 245114.
[55] T. Hörlin, T. Niklewski, M. Nygren, Mater. Res. Bull. 7 (1972) 1515.
[56] K. Shibuya, M. Kawasaki, Y. Tokura, Appl. Phys. Lett. 96 (2010) 022102.
[57] E. Sakai, K. Yoshimatsu, K. Shibuya, H. Kumigashira, E. Ikenaga, M. Kawasaki, Y. Tokura, M. Oshima, Phys. Rev. B 84 (2011) 195132.
[58] I. Balberg, B. Abeles, Y. Arie, Thin Solid Films 24 (1974) 307.
[59] M. Park, T. E. Mitchell, A. H. Heuer, J. Am. Ceram. Soc. 58 (1975) 43.
[60] Z. Hiroi, T. Yoshida, J. Yamaura, Y. Okamoto, submitted to APL Materials.
[61] Y. Muraoka, Z. Hiroi, Appl. Phys. Lett. 80 (2002) 583.
[62] B. Lazarovits, K. Kim, K. Haule, G. Kotliar, Phys. Rev. B 81 (2010) 115117.
[63] J. H. Park, J. M. Coy, T. S. Kasirga, C. Huang, Z. Fei, S. Hunter, D. H. Cobden, Nature 500 (2013) 431.
[64] J. P. Pouget, P. Lederer, D. S. Schreiber, H. Launois, D. Wohlleben, A. Casalot, G. Villeneuve, J. Phys. Chem. Solids 33 (1972) 1961.
[65] P. Lederer, H. Launois, J. P. Pouget, A. Casalot, G. Villeneuve, J. Phys. Chem. Solids 33 (1972) 1969.
[66] H. Hayamizu, Y. Muraoka, Y. Ueda, Z. Hiroi, in preparation.
[67] D. B. Rogers, R. D. Shannon, A. W. Sleight, J. L. Gillson, Inorg. Chem. 8 (1969) 841.
[68] R. D. Shannon, Solid State Commun. 6 (1968) 139.
[69] M. P. Herrero-Fernandez, B. L. Chamberland, J. Less Comm. Metals 99 (1984) 99.
[70] N. Schoenberg, Acta Chem. Scand. 8 (1954) 240.
[71] V. I. Khitrova, V. V. Klechkovskaya, Z. G. Pinsker, Kristallogr. 12 (1967) 1044.
[72] A. A. Bolzan, B. J. Kennedy, C. J. Howard, Aust. J. Chem. 48 (1995) 1473.
[73] E. E. Rodriguez, F. Poineau, A. Llobet, A. P. Sattelberger, J. Bhattacharjee, U. V. Waghmare, T. Hartmann, A. K. Cheetham, J. Am. Chem. Soc. 129 (2007) 10244.
[74] D. B. McWhan, M. Marezio, J. P. Remeika, P. D. Dernier, Phys. Rev. B 10 (1974) 490.
[75] A. K. Cheetham, C. N. R. Rao, Acta Cryst. B 32 (1976) 1579.
[76] A. Petersen, H. Müller-Buschbaum, Z. Anorg. Allg. Chem. 609 (1992) 51.
[77] R. W. Wyckoff, Crystal Structures, Interscience Publishers, New York, London, Sydney, 1965.
[78] A. Nakua, H. Yun, J. N. Reimers, J. E. Greedan, C. V. Stager, J. Solid State Chem. 91 (1991) 105.
[79] B. R. D. Shannon, C. T. Prewitt, Acta. Cryst. B25 (1969) 925.
[80] W. H. Baur, A. A. Khan, Acta Cryst. B 27 (1971) 2133.
[81] A. A. Bolzan, C. Fong, B. J. Kennedy, C. J. Howard, Acta Cryst. B 53 (1997) 373.
[82] K. Kukli, J. Aarik, A. Aidla, O. Kohan, T. Uustare, V. Sammelselg, Thin Solid Films 260 (1995) 135.
[83] M. Kitamura, K. Inoue, H. Chen, Mater. Chem. Phys. 56 (1998) 1.
[84] K. G. Bramnik, H. Ehrenberg, R. Theissmann, H. Fuess, E. Morán, Z. Kristallographie 218 (2003) 455.
[85] D. Adler, Rev. Mod. Phys. 40 (1968) 714.
[86] L. F. Mattheiss, Phys. Rev. B 13 (1976) 2433.
[87] A. F. Wells, Structural Inorganic Chemistry, Clarendon Press, Oxford, 1984.
[88] Y. Tsuchiya, T. Tagi, Nature 347 (1990) 267.
[89] W. Baur, Acta Cryst. 9 (1956) 515.
[90] W. H. Baur, Acta Cryst. B 32 (1976) 2200.
[91] A. A. Bolzan, C. Fong, B. J. Kennedy, C. J. Howard, J. Solid State Chem. 113 (1994) 9.
[92] K. Sugiyama, Y. Takéuchi, Z. Kristallographie 194 (1991) 305.
[93] C. J. Howard, B. J. Kennedy, C. Curfs, Phys. Rev. B 72 (2005) 214114.
[94] J. R. Brews, Phys. Rev. B 1 (1970) 2557.
[95] F. Gervais, W. Kress, Phys. Rev. B 31 (1985) 4809.
[96] R. Srivastava, L. L. Chase, Phys. Rev. Lett. 27 (1971) 727.
[97] S. M. Shapiro, J. D. Axe, G. Shirane, P. M. Raccah, Solid State Commun. 15 (1974) 377.
[98] F. A. Cotton, Quart. Rev. 20 (1966) 389.
[99] T. P. Sleight, C. R. Hare, A. W. Sleight, Mater. Res. Bull. 3 (1968) 437.
[100] J. P. Pouget, H. Launois, J. P. D'Haenens, P. Merenda, T. M. Rice, Phys. Rev. Lett. 35 (1975) 873.
[101] U. Müller, Inorganic Structural Chemistry, John Wiley & Sons, Ltd., 2007.
[102] A. W. Sleight, Prog. Solid State Chem. 37 (2009) 251.
[103] S. Merlino, L. Labella, F. Marchetti, S. Toscani, Chem. Mater. 16 (2004) 3895.
[104] D. R. Taylor, J. C. Calabrese, E. M. Larsen, Inorg. Chem. 16 (1977) 721.
[105] F. A. Cotton, C. E. Rice, Inorg. Chem. 16 (1977) 1865.
[106] D. B. McWhan, J. P. Remeika, Phys. Rev. B 2 (1970) 3734.
[107] P. D. Dernier, M. Marezio, Phys. Rev. B 2 (1970) 3771.
[108] D. Adler, H. Brooks, Phys. Rev. 155 (1967) 826.
[109] S. Shin, S. Suga, M. Taniguchi, M. Fujisawa, H. Kanzaki, A. Fujimori, H. Daimon, Y. Ueda, K. Kosuge, S. Kachi, Phys. Rev. B 41 (1990) 4993.
[110] A. Bianconi, Phys. Rev. B 26 (1982) 2741.
[111] C. Sommers, R. Degroot, D. Kaplan, A. Zylbersztejn, J. Phys. Lett. 36 (1975) L157.
[112] D. B. McWhan, J. P. Remeika, J. P. Maita, H. Okinaka, K. Kosuge, S. Kachi, Phys. Rev. B 7 (1973) 326.
[113] P. F. Bongers, Solid State Commun. 3 (1965) 275.

[114] J. L. Hodeau, M. Marezio, C. Roucau, R. Ayroles, A. Meerschaut, J. Rouxel, P. Monceau, J. Phys. C 11 (1978) 4117.
[115] A. Cavalleri, T. Dekorsy, H. H. W. Chong, J. C. Kieffer, R. W. Schoenlein, Phys. Rev. B 70 (2004) 161102.
[116] P. Baum, D.-S. Yang, A. H. Zewail, Science 318 (2007) 788.
[117] K. Shibuya, M. Kawasaki, Y. Tokura, Phys. Rev. B 82 (2010) 205118.
[118] J. B. Goodenough, Buul. Soc. Chim. Fr. 4 (1961) 1200.
[119] T. Kawakubo, J. Phys. Soc. Jpn. 20 (1965) 516.
[120] J. C. Slater, Phys. Rev. 82 (1951) 538.
[121] G. Pourroy, M. Drillon, L. Padel, J. C. Bernier, Physica B+C 123 (1983) 21.
[122] R. R. Neurgaonkar, R. Roy, Inorg. Chem. 16 (1977) 3349.
[123] P. Khalifah, R. J. Cava, Phys. Rev. B 64 (2001) 085111.
[124] Y. Miura, Y. Yasui, M. Sato, N. Igawa, K. Kakurai, J. Phys. Soc. Jpn. 76 (2007) 033705.
[125] W. H. McCarroll, L. Katz, R. Ward, J. Am. Chem. Soc. 79 (1957) 5410.
[126] W. H. McCarroll, Inorg. Chem. 16 (1977) 3351.
[127] C. C. Torardi, R. E. McCarley, Inorg. Chem. 24 (1985) 476.
[128] S. J. Hibble, I. D. Fawcett, Inorg. Chem. 34 (1995) 500.
[129] J. P. Sheckelton, J. R. Neilson, D. G. Soltan, T. M. McQueen, Nat Mater 11 (2012) 493.
[130] C. C. Torardi, J. C. Calabrese, Inorg. Chem. 23 (1984) 3281.
[131] T. Toriyama, M. Watanabe, T. Konishi, Y. Ohta, Phys. Rev. B 88 (2013) 235116.
[132] T. Ozawa, I. Suzuki, H. Sato, J. Phys. Soc. Jpn. 75 (2006) 014802.
[133] Y. Miura, M. Sato, Y. Yamakawa, T. Habaguchi, Y. Ōno, J. Phys. Soc. Jpn. 78 (2009) 094706.
[134] G. Jackeli, D. I. Khomskii, Phys. Rev. Lett. 100 (2008) 147203.
[135] J. B. Goodenough, G. Dutta, A. Manthiram, Phys. Rev. B 43 (1991) 10170.
[136] N. Katayama, M. Uchida, D. Hashizume, S. Niitaka, J. Matsuno, D. Matsumura, Y. Nishihata, J. Mizuki, N. Takeshita, A. Gauzzi, M. Nohara, H. Takagi, Phys. Rev. Lett. 103 (2009) 146405.
[137] C. A. Bridges, J. E. Greedan, J. Solid State Chem. 177 (2004) 1098.
[138] T. Kajita, T. Kanzaki, T. Suzuki, J. E. Kim, K. Kato, M. Takata, T. Katsufuji, Phys. Rev. B 81 (2010) 060405.
[139] M. Isobe, S. Koishi, N. Kouno, J. Yamaura, T. Yamauchi, H. Ueda, H. Gotou, T. Yagi, Y. Ueda, J. Phys. Soc. Jpn. 75 (2006) 073801.
[140] M. Isobe, E. Ninomiya, A. N. Vasil'ev, Y. Ueda, J. Phys. Soc. Jpn. 71 (2002) 1423.
[141] E. Ninomiya, M. Isobe, Y. Ueda, M. Nishi, K. Ohoyama, H. Sawa, T. Ohama, Physica B 329–333 (2003) 884.
[142] M. J. Konstantinović, J. van den Brink, Z. V. Popović, V. V. Moshchalkov, M. Isobe, Y. Ueda, Phys. Rev. B 69 (2004) 020409.
[143] T. Hikihara, Y. Motome, Phys. Rev. B 70 (2004) 214404.
[144] M. Isobe, Y. Ueda, J. Phys. Soc. Jpn. 71 (2002) 1848.
[145] Y. Okamoto, S. Niitaka, M. Uchida, T. Waki, M. Takigawa, Y. Nakatsu, A. Sekiyama, S. Suga, R. Arita, H. Takagi, Phys. Rev. Lett. 101 (2008) 086404.
[146] Y. Horibe, M. Shingu, K. Kurushima, H. Ishibashi, N. Ikeda, K. Kato, Y. Motome, N. Furukawa, S. Mori, T. Katsufuji, Phys. Rev. Lett. 96 (2006) 086406.
[147] P. G. Radaelli, Y. Horibe, M. J. Gutmann, H. Ishibashi, C. H. Chen, R. M. Ibberson, Y. Koyama, Y.-S. Hor, V. Kiryukhin, S.-W. Cheong, Nature 416 (2002) 155.
[148] A. Irizawa, S. Suga, G. Isoyama, K. Shimai, K. Sato, K. Iizuka, T. Nanba, A. Higashiya, S. Niitaka, H. Takagi, Phys. Rev. B 84 (2011) 235116.
[149] L. Pinsard-Gaudart, N. Dragoe, P. Lagarde, A. Flank, J. Itie, A. Congeduti, P. Roy, S. Niitaka, H. Takagi, Phys. Rev. B 76 (2007) 045119.
[150] D. I. Khomskii, T. Mizokawa, Phys. Rev. Lett. 94 (2005) 156402.
[151] I. I. Mazin, H. O. Jeschke, K. Foyevtsova, R. Valentí, D. I. Khomskii, Phys. Rev. Lett. 109 (2012) 197201.
[152] M. S. Senn, J. P. Wright, J. P. Attfield, Nature 481 (2012) 173.
[153] M. S. Senn, I. Loa, J. P. Wright, J. P. Attfield, Phys. Rev. B 85 (2012) 125119.
[154] J. P. Attfield, APL Materials 3 (2015) 041510.
[155] S. H. Lee, D. Louca, H. Ueda, S. Park, T. J. Sato, M. Isobe, Y. Ueda, S. Rosenkranz, P. Zschack, J. Íñiguez, Y. Qiu, R. Osborn, Phys. Rev. Lett. 93 (2004) 156407.
[156] H. Tsunetsugu, Y. Motome, Phys. Rev. B 68 (2003) 060405.
[157] S. H. Lee, C. Broholm, T. H. Kim, W. Ratcliff, S. W. Cheong, Phys. Rev. Lett. 84 (2000) 3718.
[158] K. Takanashi, H. Yasuoka, Y. Ueda, K. Kosuge, J. Phys. Soc. Jpn. 52 (1983) 3953.
[159] C. Schlenker, S. Lakkis, J. M. D. Coey, M. Marezio, Phys. Rev. Lett. 32 (1974) 1318.
[160] P. Porta, M. Marezio, J. P. Remeika, P. D. Dernier, Mater. Res. Bull. 7 (1972) 157.
[161] J. Haines, J. M. Léger, O. Schulte, S. Hull, Acta Cryst. B 53 (1997) 880.
[162] C. E. Boman, Acta Chem. Scand. 24 (1970) 123.
[163] P. D'Antonio, A. Santoro, Acta Cryst. B 36 (1980) 2394.
[164] B. Jasper-Tönnies, H. Müller-Buschbaum, Z. Anorg. Allg. Chem. 504 (1983) 113.
[165] L. W. Vernon, W. O. Milligan, Texas Journal of Science 1 (1951) 82.
[166] J. Isasi, M. L. López, M. L. Veiga, C. Pico, J. Alloys Comp. 232 (1996) 36.
[167] D. N. Astrov, N. A. Kryukova, R. B. Zorin, V. A. Makarov, R. P. Ozerov, F. A. Rozhdestvenskii, V. P. Smirnov, A. M. Turchaninov, N. V. Fadeeva, Kristallogr. 17 (1972) 1152.
[168] J. Amador, I. Rasines, J. Appl. Cryst. 14 (1981) 348.
[169] Ismunandar, B. J. Kennedy, B. A. Hunter, Mater. Sci. Forum 278 (1998) 714.

**Table 2**
Crystal data and the electronic property of the $AO_2$ compounds having the tetragonal rutile structure.

| Compound | $a$ (Å) | $c$ (Å) | $V$ (Å$^3$) | $c/a$ | $x$(O) | Electron config. | Electronic property[a] | Reference |
|---|---|---|---|---|---|---|---|---|
| $SiO_2$ | 4.179 | 2.6649 | 46.5399 | 0.6377 | 0.3062 | $2p^6$ | I | [80] |
| $TiO_2$ | 4.5941 | 2.9589 | 62.4498 | 0.6441 | 0.3057 | $3d^0$ | I | [80] |
| $VO_2$[b] | 4.5546 | 2.8514 | 59.1796 | 0.6264 | 0.3001 | $3d^1$ | M | [74] |
| $CrO_2$ | 4.419 | 2.9154 | 56.9307 | 0.6597 | 0.3026 | $3d^2$ | M | [160] |
| $MnO_2$ | 4.3983 | 2.873 | 55.5783 | 0.6532 | 0.30515 | $3d^3$ | I | [90] |
| $GeO_2$ | 4.40656 | 2.86186 | 55.5709 | 0.6495 | 0.30604 | $3d^{10}$ | I | [81] |
| $NbO_2$[c] | 4.8463 | 3.0315 | 71.1997 | 0.6255 | 0.2924 | $4d^1$ | M | [91] |
| $RuO_2$ | 4.4968 | 3.1049 | 62.7108 | 0.6905 | 0.3053 | $4d^4$ | M | [81,161] |
| $RhO_2$ | 4.4862 | 3.0884 | 62.1571 | 0.6884 | 0.318 | $4d^5$ | M | [68] |
| $SnO_2$ | 4.73735 | 3.1864 | 71.5107 | 0.6726 | 0.30562 | $4d^{10}$ | I | [81] |
| $TaO_2$ | 4.709 | 3.065 | 67.9654 | 0.6509 | 0.303 | $5d^1$ | ? | [70] |
| $TeO_2$[d] | 4.79 | 3.77 | 86.50 | 0.7875 | - | $5s^2$ | I | [77] |
| $OsO_2$ | 4.5003 | 3.1839 | 64.4826 | 0.7075 | 0.3081 | $5d^4$ | M | [162] |
| $IrO_2$ | 4.5051 | 3.1586 | 64.1067 | 0.7011 | 0.3077 | $5d^5$ | M | [81] |
| β'-$PtO_2$[e] | 4.485 | 3.130 | 62.9607 | 0.6979 | 0.305 | $5d^6$ | ? | [69] |
| β-$PbO_2$ | 4.9578 | 3.3878 | 83.2714 | 0.6833 | 0.3067 | $5d^{10}$ | I | [163] |

[a]Insulator (I) or metal (M) mostly from Rogers et al. [67]; [b]obtained at 360 K; [c]obtained at 1273 K; [d]polymorph in the rutile structure but not considered in the present paper; [e]prepared at high pressures of 4–6 GPa and 723 K; $PtO_2$ may exist in three modifications: α ($CdI_2$ type), β ($CaCl_2$ type), and β' (rutile type) [68].

**Table 3**
Crystal data and the electronic property of the $AO_2$ compounds having distorted rutile structures. The corresponding lattice parameters in the regular rutile structure are also given.

| Compound | $a$ (Å) | $b$ (Å) | $c$ (Å) | $β$ (°) | $a_r$ (Å) | $c_r$ (Å) | $V_r$ (Å$^3$) | $c_r/a_r$ | Electron config. | Electronic property | Reference |
|---|---|---|---|---|---|---|---|---|---|---|---|
| $VO_2$[a] | 5.7515 | 4.5252 | 5.3819 | 122.6 | 4.5296 | 2.87575 | 59.0024 | 0.6349 | $3d^1$ | I | [67] |
| $NbO_2$[b] | 13.696 | | 5.981 | | 4.8423 | 2.9905 | 70.1199 | 0.6176 | $4d^1$ | I | [75,91] |
| $MoO_2$ | 5.6102 | 4.8573 | 5.6265 | 120.915 | 4.8422 | 2.8051 | 65.7709 | 0.5793 | $4d^2$ | M | [72] |
| $TcO_2$ | 5.6891 | 4.7546 | 5.5195 | 121.453 | 4.7316 | 2.84455 | 63.6812 | 0.6012 | $4d^3$ | M? | [73] |
| $WO_2$ | 5.5769 | 4.8986 | 5.6644 | 120.678 | 4.8851 | 2.78845 | 66.5444 | 0.5708 | $5d^2$ | M | [72] |
| α-$ReO_2$ | 5.562 | 4.838 | 5.561 | 120.87 | 4.8056 | 2.781 | 64.2208 | 0.5787 | $5d^3$ | M | [67] |

[a]obtained at 298 K; [b]obtained at 295 K, space group $I4_1/a$

**Table 4**
Crystal data for the $AA'O_4$ compounds having the rutile structure.

| Compound | $a$ (Å) | $c$ (Å) | $V$ (Å$^3$) | $c/a$ | $x$(O) | Reference |
|---|---|---|---|---|---|---|
| $AlSbO_4$ | 4.51 | 2.961 | 60.2270 | 0.6565 | | [77] |
| $AlTaO_4$ | 4.6065 | 2.985 | 63.3412 | 0.6480 | 0.302 | [164] |
| $TiVO_4$ | 4.58 | 2.95 | 61.8804 | 0.6441 | 0.3 | [165] |
| $TiNbO_4$ | 4.743 | 2.9944 | 67.3622 | 0.6313 | 0.295 | [76] |
| $TiTaO_4$ | 4.709 | 3.0672 | 68.0142 | 0.6513 | 0.3 | [76] |
| $VRhO_4$ | 4.567 | 2.924 | 60.9873 | 0.6402 | 0.302 | [166] |

| Compound | | | | | | | |
|---|---|---|---|---|---|---|---|
| VSbO$_4$ | 4.58 | 3.06 | 64.1878 | 0.6681 | 0.3 | | [165] |
| VTaO$_4$ | 4.667 | 3.047 | 66.3664 | 0.6529 | | | [167] |
| VReO$_4$ | 4.6357 | 2.8292 | 60.7987 | 0.6103 | 0.294 | | [84] |
| CrNbO$_4$ | 4.6484 | 3.0113 | 65.0670 | 0.6478 | 0.303 | | [76] |
| CrSbO$_4$ | 4.577 | 3.042 | 63.7266 | 0.6646 | | | [77] |
| CrTaO$_4$ | 4.6445 | 3.0186 | 65.1154 | 0.6499 | 0.304 | | [76] |
| FeNbO$_4$ | 4.68 | 3.05 | 66.8023 | 0.6517 | | | [77] |
| FeSbO$_4$ | 4.6388 | 3.0773 | 66.2188 | 0.6634 | 0.315 | | [168] |
| FeTaO$_4$ | 4.679 | 3.047 | 66.7081 | 0.6512 | | | [167] |
| GaSbO$_4$ | 4.59 | 3.03 | 63.8363 | 0.6601 | | | [77] |
| RhNbO$_4$ | 4.70186 | 3.01685 | 66.6950 | 0.6416 | 0.299 | | [169] |
| RhSbO$_4$ | 4.62189 | 3.10721 | 66.3758 | 0.6723 | 0.307 | | [169] |
| RhTaO$_4$ | 4.6971 | 3.0303 | 66.8567 | 0.6451 | 0.302 | | [169] |

**Table 5**
Candidates for the molecular orbital crystal and their electronic and magnetic properties. Each compound shows a metal-insulator transition (MIT) or an insulator-insulator transition (IIT) at $T_{MI}$, or remain a metal (M) or an insulator (I) at all temperatures. The magnetic property at the lowest temperatures is classified into four groups; nonmagnetic/diamagnetic (N) and paramagnetic (P) below $T_{MI}$, or antiferromagnetic (AF) and ferrimagnetic (F) below $T_{mag}$.

| Compound | Molecule | Structure type | Electron config. | Electronic property | Magnetic property | $T_{MI}$, $T_{mag}$ (K) | Related compounds | Reference |
|---|---|---|---|---|---|---|---|---|
| VO$_2$ | dimer | rutile | $3d^1$ | MIT | N | 340 | – | [67] |
| NbO$_2$ | dimer | rutile | $4d^1$ | MIT | N | 1070 | – | [75,91] |
| MoO$_2$ | dimer | rutile | $4d^2$ | M | P | – | TcO$_2$, WO$_2$, α-ReO$_2$ | [67,72] |
| Ti$_4$O$_7$ | dimer | Magnéli phase | $3d^{0.5}$ | MIT | N | 149 | Ti$_n$O$_{2n-1}$ ($n$ = 3, 5, 6) | [34,36-38] |
| V$_4$O$_7$ | dimer | Magnéli phase | $3d^{1.5}$ | MIT | AF | 250, 40 | V$_n$O$_{2n-1}$ ($n$ = 6, 8, 9) | [36,39,40] |
| NbCl$_4$ | dimer | chain (edge) | $4d^1$ | I | N | – | NbI$_4$, WCl$_4$ | [104] |
| MoBr$_3$ | dimer | chain (face) | $4d^3$ | I | N | – | β-TiCl$_3$, ZrI$_3$, RuCl$_3$, RuBr$_3$ | [103] |
| NaTiSi$_2$O$_6$ | dimer | pyroxene | $3d^1$ | IIT | N | 210 | LiTiSi$_2$O$_6$ | [140] |
| K$_2$V$_8$O$_{16}$ | dimer | hollandite | $3d^{1.25}$ | MIT | N | 170 | – | [139] |
| MgTi$_2$O$_4$ | dimer | spinel | $3d^1$ | IIT | N | 260 | – | [144] |
| LiRh$_2$O$_4$ | dimer | spinel | $4d^{5.5}$ | MIT | N | 170 | – | [145] |
| Li$_2$RuO$_3$ | dimer | honeycomb | $4d^4$ | IIT | N | 540 | – | [124,133] |
| Fe$_3$O$_4$ | trimer | spinel | $3d^{5.33}$ | MIT | F | 120, 858 | – | [152,153] |
| LiVO$_2$ | trimer | triangular | $3d^2$ | MIT | N | 490 | LiVS$_2$ | [135] |
| BaV$_{10}$O$_{15}$ | trimer | complex | $3d^{2.2}$ | MIT | P | 130 | – | [137,138] |
| Zn$_2$Mo$_3$O$_8$ | trimer | kagome | $4d^2$ | I | N | – | – | [125] |
| LiZn$_2$Mo$_3$O$_8$ | trimer | kagome | $4d^{2.33}$ | I | P | – | – | [127,129] |
| K$_2$Mo$_8$O$_{16}$ | tetramer | hollandite | $4d^{2.25}$ | I | P | – | Rb$_{1.5}$Mo$_8$O$_{16}$ | [130,132] |
| AlV$_2$O$_4$ | heptamer | spinel | $3d^{2.5}$ | MIT | P | 700 | – | [146] |

| CuIr$_2$S$_4$ | octamer (four dimers) | spinel | 5$d^{5.5}$ | MIT | N | 230 | – | [147] |
| Na$_2$IrO$_3$ | hexamer | honeycomb | 5$d^5$ | I | P | – | – | [151] |